\documentclass[10pt,a4paper]{article}
\usepackage{graphicx}
\usepackage{amsfonts, amsbsy, amssymb, amsmath, graphicx, float, subfigure, color}
\usepackage{hyperref}
\newsavebox{\astrutbox}
\sbox{\astrutbox}{\rule[-5pt]{0pt}{20pt}}

\def\Ra{\mbox{\textrm{Ra}}}
\newcommand\Pran{\mbox{\textrm{Pr}}} % Prandtl number, cf TeX's \Pr product

\usepackage{amsmath,amssymb,graphicx}
\usepackage{stfloats}
\usepackage{subfigure}

\begin{document}

%\title[Dynamics of  convection with temperature-dependent viscosity with  O(2) symmetry.]
\title{Bifurcations and dynamics
in convection with temperature-dependent viscosity in the presence of the O(2) symmetry }

%\author[J. Curbelo and A.M. Mancho]%
%{J\ls E\ls Z\ls A\ls B\ls E\ls L\ls  \ns C\ls U\ls R\ls B\ls E\ls L\ls O$^{1,2}$%
%  ,\ns
%H.\ls-\ls C.\ns S\ls M\ls I\ls T\ls H$^1$\break
%\and A\ls N\ls A\ns    M.\ns  M\ls A\ls N\ls C\ls H\ls O$^1$}
%\thanks{Correspondence to: a.m.mancho@icmat.es.}
\author{J. Curbelo$^{1,2}$, A. M. Mancho$^1$\\
$^1$Instituto de Ciencias Matem\'aticas (CSIC-UAM-UCM-UC3M),\\ Nicol\'as Cabrera, 13-15,
28049 Madrid, Spain\\
$^2$Departamento de Matem\'aticas, Facultad de Ciencias, \\Universidad Aut\'onoma de Madrid,
28049 Madrid, Spain}
\maketitle

%\email{jezabel.curbelo@icmat.es, a.m.mancho@icmat.es}


\maketitle

\begin{abstract}
{We focus  the study of a convection problem in a 2D set--up in the presence of the O(2) symmetry.
The viscosity in the fluid depends on the temperature as it  changes its value abruptly in an interval around a temperature of transition.
The influence of the viscosity law on the  morphology of the plumes is examined for several parameter settings, and   a variety of shapes ranging from spout  to mushroom shaped is found.
We  explore the impact of the symmetry on the time evolution of this type of fluid, and find solutions which are greatly influenced by its presence: at a large aspect ratio and high Rayleigh numbers,  traveling waves, heteroclinic connections and chaotic regimes are found. These solutions, which are due to the symmetry presence,  have not been previously described in the context of  temperature dependent viscosities. However,
similarities are found with solutions described in other contexts such as flame propagation problems or convection problems with constant viscosity also under  the presence of the O(2) symmetry, thus confirming the determining role of the symmetry  in the dynamics.}
%This could suggest  that  abrupt changes in the viscosity could define a front similar to those observed in  flame propagation phenomena. }

\end{abstract}

%\begin{keyword}
%spectral semi-implicit method, numerical analysis, convection with viscosity dependent on temperature
%%% keywords here, in the form: keyword \sep keyword

%35-04, 65-04, 65M70, 76R05, 76-04
%%%MSC codes here, in the form: \MSC code \sep code
%%% or \MSC[2008] code \sep code (2000 is the default)

%\end{keyword}

% \linenumbers

\section{Introduction}
\label{S1}
{This paper addresses the numerical study  of convection at  infinite Prandtl number in fluids in which viscosity strongly depends on temperature   in the presence of O(2) symmetry.
Convection in fluids with temperature-dependent viscosity is of interest because of its importance in engineering and geophysics.
 Linear and quadratic dependencies of the viscosity on temperature   have been discussed \cite{PEG67,RS93, DS04,VW06}, but in order  to address the Earth's upper mantle convection, in which viscosity contrasts are
  of several orders of magnitude, a stronger dependence with temperature must be considered.  This problem   has been approached both in experiments \cite{RND83,CG94,Boo76,W88} and in theory \cite{OSZ91,BMM92,RTSZ97,MS95,S12,SB07}.  In these contexts, the dependence of  viscosity with temperature is expressed by means of an Arrhenius law.
In \cite{BMM92}, the exponential dependence is discussed
as an approach to the Arrhenius law by means of a Taylor
series around a reference temperature. This is also called the
Frank-Kamenetskii approximation (see \cite{FB02}). Viscosity has also been considered when it depends  on other magnitudes
such as depth  \cite{BRB96,BBCCGHHJKMMOSS89},  a combination of both depth and temperature \cite{BBCCGHHJKMMOSS89}
or pressure \cite{RePa87}. However, it is commonly accepted \cite{Dav01,RePa87} that in the Earth's interior, viscosity depends
most significantly on temperature. The usual approach in numerical models of the mantle \cite{MS95,BBCCGHHJKMMOSS89,TT71}  is to consider constant thermal conductivity. This approach has also been verified in fluid experiments seeking to model
mantle convection \cite{Boo76}. However,  studies also exist which consider variations on thermal conductivity \cite{DYR99,DYY00,YNY04}.

Here, we focus on the study of a fluid in which  the viscosity changes abruptly in a temperature interval around a temperature of transition. This defines a phase
change over  a mushy region, which expresses   the melting of minerals or other components.  Melting and solidification processes are  important  in magma chamber dynamics \cite{BJ86,BM89}, in volcanic conduits \cite{DB05,M02}, in the formation of chimneys in mushy layers \cite{CGJH70}, in  metal processing in industry (see, for example, \cite{SH88}), etc.
In  phase transitions,  other fluid properties in addition to viscosity may change abruptly, such as  density or  thermal diffusivity. However, in this study we consider solely the study of  effects due to the variability of viscosity,
since consideration of the effect of  simultaneous variations on all the properties prevents a focused  understanding of the exact role played by each one of these properties.
Viscosity  is a measure of  fluid  resistance to gradual deformation, and in this sense very viscous fluids are more likely to behave rigidly when compared to less viscous fluids. When examining the proposed transition   with temperature, we focus on the global fluid motion when some parts of it tend to be more rigid than others. Disregarding  the   variations on density in this transition moves us away from  instabilities  caused by abrupt density changes such as the Rayleigh Taylor instability, in which a denser fluid over a lighter one tends to penetrate it by forming a fingering pattern. A recent article  by M. Ulvrov\'a  {\it et al.} \cite{ULCRT12} deals with a similar problem to ours, but  takes into account both variations in density and  in viscosity. Thermal conductivity  effects  are related to the relative importance of heat advection versus diffusion. In this way,  diffusive effects
are important at large conductivity, while  heat advection by fluid particles is dominant at low conductivity. The contrasts  arising from these variations are beyond the scope of our work and thus are disregarded here.

This paper addresses the convection of a 2D fluid layer with  temperature-dependent viscosity  and periodic boundary conditions {\it possessing the O(2) symmetry}.
The motivation arises from the fact that symmetric systems typically exhibit more complicated bifurcations than non-symmetric systems and introduce  conditions and degeneracies in bifurcation analysis. There exist numerous novel dynamical phenomena whose existence is fundamentally related to the presence of symmetry \cite{CK91,GS85,GSS00,F80}.  Solutions related
to the presence of symmetry, include rotating waves \cite{R73}, modulated waves \cite{R82,AGH88}, slow ``phase" drifts along
directions of broken symmetry \cite{K90}, and stable heteroclinic cycles \cite{AGH88,GH88,PoMa97}.
The SO(2) symmetry is present   in  problems  described by  the Navier-Stokes \cite{GS86,HS90} or the Kuramoto-Sivashinsky \cite{PoMa97,AGH89} equations with periodic boundary conditions, since the equations are invariant  under translations  and  the boundary conditions do not break this invariance.  Additionally,  if  the reflection symmetry exists, %along the translational direction,
the full symmetry group is the O(2) group.  While in classical convection problems (with constant viscosity),
the study of the solutions and bifurcations in the presence of symmetries has been the object of much attention \cite{KBS88,GSK84,OMMG93,MOMGG94,KMG96,EKG97,DCJ00,ABK07},  its counterpart in fluids  with viscosity depending on temperature has received less consideration. Our 2D physical set-up is idealized with respect to realistic geophysical flows occurring in the Earth's interior, as these are 3D flows moving in spherical shells \cite{B75,B82}. Under these conditions,
the symmetry present in the problem  is formed by all the orientation preserving rigid motions of $ \mathbb{R}^3$ that fix the origin,which is the SO(3) group \cite{C75,GS82,IG84}. The effects of
the  Earth's  rotation are negligible in this respect and do not break this symmetry, since the high viscosity of the mantle makes the   Coriolis number insignificant. The link between our simplified problem  and these realistic set-ups is that  the O(2) symmetry is isomorphic   to the rotations along the azimuthal coordinate, which form a closed subgroup of SO(3). Additionally, the O(2) symmetry is present in systems with cylindrical geometry, which provide an idealized setting for volcanic conduits and magma chambers. The SO(2)  symmetry is also present in 3D flows  moving in spherical shells which rotate around an axis.

In this way,  specific symmetry-related solutions found in our setting are expected  to be present in these other contexts.
The interest of 2D numerical studies for representing 3D time-dependent thermal convection with constant viscosity has been addressed in \cite{SBH04}. The authors report that  in turbulent regimes  at  high Rayleigh numbers,
 the flow structure and global quantities such as the Nusselt number  and the Reynolds number show a similar behaviour in 3D and 2D simulations as far as high values of the Prandtl number
 are concerned. In some sense, these results suggest that our simulations might be illustrative for the 3D case,  since  although  they are far from a turbulent regime and do not correspond to the case of  constant viscosity,   they have been performed according to the infinite Prandtl number approach. In this article  we show  that typical solutions  of systems with symmetries, as previously reported in diverse contexts  \cite{AGH88,PoMa97,vGMP86}, are also present in mantle convection and magma-related problems. We report the presence of traveling waves and  limit cycles near heteroclinic connections after a Hopf bifurcation. We do this by means of bifurcation analysis techniques an direct numerical simulations and  of the full partial differential equations system.}
 %Despite our simplifications what is  of interest about the O(2) symmetry present in our study, and that links it to more realistic set-ups, is that it is isomorphic   to the rotations along the azimuthal coordinate which is a closed subgroup of SO(3).

The article is organized as follows: In Section \ref{S2}, we formulate the problem, providing the description of the physical set-up, the basic equations and boundary conditions.
In Section \ref{Svis} we present the  viscosity law under consideration and discuss several limits in which previously studied dependencies are recovered.
Section \ref{SSolutions} summarizes the numerical methods used to   sketch an outlook of the solutions displayed by the system.
Section \ref{Sresults}    discusses  the   solutions obtained for a broad  parameter set.  Finally Section \ref{S5} presents the conclusions.


 \section{Formulation of the problem}\label{S2}

\begin{figure}
     \centering
   \includegraphics[width=8.5cm]{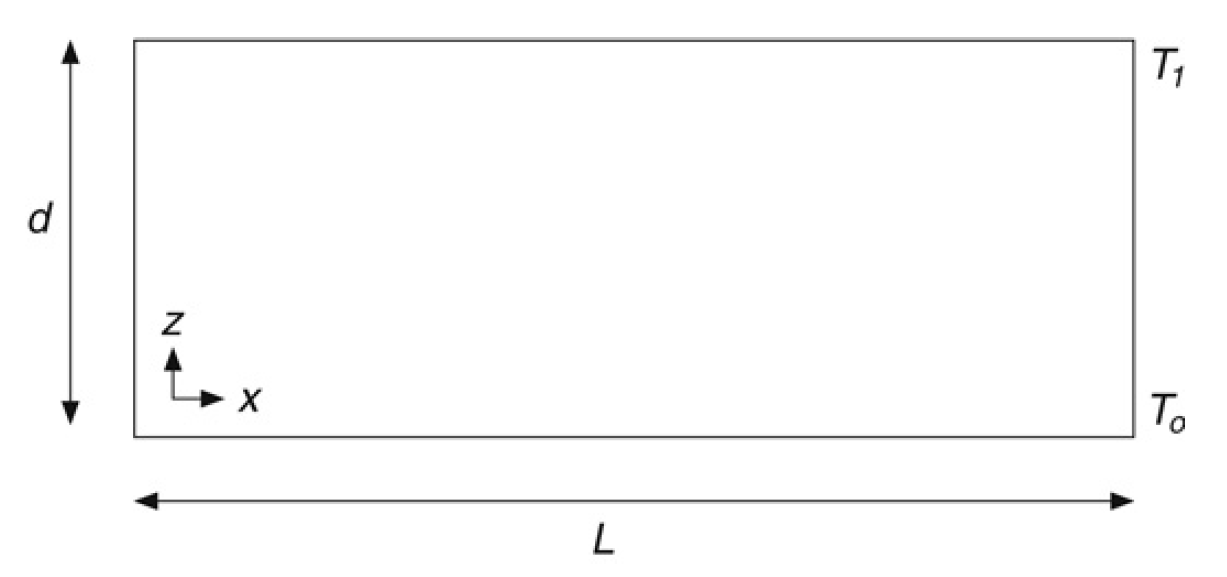}
    \caption{Problem set-up. {A 2D container of length $L$ and depth $d$ with periodic lateral boundary conditions. The bottom plate plate is at temperature $T_0$ and is rigid, while at the upper plate the temperature is  $T_1$ ($T_1<T_0$) and free slip. }}\label{F1}
   \end{figure}

   As shown in Figure \ref{F1} we consider a fluid layer, placed in a  2D container  of length $L$  ($x$ coordinate) and depth $d$ ($z$ coordinate).    The bottom plate is at temperature $T_0$  and  the upper plate  is at $T_1,$ where $T_1=T_0-\Delta T$ and $\Delta T$ is the vertical temperature difference, which is positive, {\it i.e},  $T_1<T_0$.

 The magnitudes involved in the equations governing the system are the velocity field  $\mathbf{u}=(u_x,u_z)$, the temperature $T$, and the pressure $P$. The spatial coordinates are  $x$ and $z$  and the time is denoted by $t$. Equations are simplified by taking into account the Boussinesq approximation, where the density $\rho$ is considered as constant everywhere except in the external forcing term, where a dependence on temperature is assumed, as follows $\rho=\rho_0(1-\alpha(T-T_1)).$ Here $\rho_0$ is the mean density at temperature $T_1$ and $\alpha$ the thermal expansion coefficient.

 The equations  are expressed with magnitudes in dimensionless form after
rescaling as  follows: $(x', z') = (x,z)/ d$, $t' = \kappa t / d^2$,
$\mathbf{u}'=  d \mathbf{u}/\kappa$, $P' = d^2 P / (\rho_0 \kappa \nu_0)$ , $\theta'= (T - T_1) /(\Delta T)$.
Here, $\kappa$ is the thermal diffusivity and $\nu_0$ is the maximum viscosity of the fluid, which is viscosity at temperature $T_1$.
After rescaling the domain, $\Omega_1 = [0,L)\times[0,d]$
is transformed into $\Omega_2 = [0 ,\Gamma)\times[  0 , 1] $ where
$\Gamma  = L / d$ is the aspect ratio.   The system evolves according to the momentum and the
mass balance equations, as well a to the energy conservation principle.  The non-dimensional equations  are (after dropping the primes in the fields):
   \begin{align}
&\nabla \cdot \mathbf{u}=0, \label{eqproblem1}\\
&\frac{1}{\Pran}(\partial_t \mathbf{u}+ \mathbf{u}\cdot \nabla \mathbf{u}) = \Ra\theta\vec{e}_3-\nabla P+\text{div} \left(\frac{\nu(\theta)}{\nu_0}(\nabla \mathbf{u}+(\nabla \mathbf{u})^T)\right),\label{eqproblem2}\\
&\partial_t \theta + \mathbf{u} \cdot \nabla \theta= \Delta \theta.\label{eqproblem3}
\end{align}

Here, $\vec{e}_3$ represents the unitary vector in the vertical direction, $\Ra=d^4 \alpha g \Delta T/(\nu_0 \kappa)$ is the Rayleigh number, $g$ is the gravity acceleration, $\Pran=\nu_0/\kappa$ is the Prandtl number. Typically for rocks $\Pran$,  is very large, since they present low thermal conductivity (approximately $10^{-6}m^2/s$) and very large viscosity (of the order $10^{20} N s/m^2$) \cite{Dav01}. Thus, for the problem under consideration, $\Pran$ can be considered as infinite and the left-hand side term in \eqref{eqproblem2} can be made equal to zero. The viscosity $\nu(\theta)$ is a smooth positive  bounded function of $\theta$, which in our set-up
 represents a transition in the fluid, due for instance  to the melting of minerals caused by an abrupt change in viscosity at a certain temperature. This is discussed in detail in the following section.

For the boundary conditions, we consider that  the bottom plate is rigid and that the upper surface is non-deformable and free slip.
The dimensionless boundary conditions are expressed as,
\begin{align}\label{eqbc}
\theta=1, \ \mathbf{u}=\vec{0},\text{ on } z=0 \text{ and } \
\theta=\partial_z u_x=u_z= 0,\text{ on } z=1.
\end{align}
Lateral  boundary conditions are periodic.  Jointly with equations (\ref{eqproblem1})-(\ref{eqproblem3}), these conditions are invariant under translations along the $x$-coordinate, which introduces the symmetry SO(2) into the problem. In convection problems with constant viscosity, the reflexion symmetry $x\to -x$ is also present   insofar as  the fields are conveniently transformed as follows $(\theta,u_x,u_z,p) \to(\theta,-u_x,u_z,p)$. In this  case,   the O(2)  group expresses the full problem symmetry.
  The new terms introduced by the temperature dependent viscosity, in the current set-up {equation} (\ref{eqproblem2})  maintain the reflexion symmetry, and the symmetry group is O(2).

\section{The viscosity law}
\label{Svis}
\begin{figure}
\centering
	    \subfigure[]{\includegraphics[width=4.5cm]{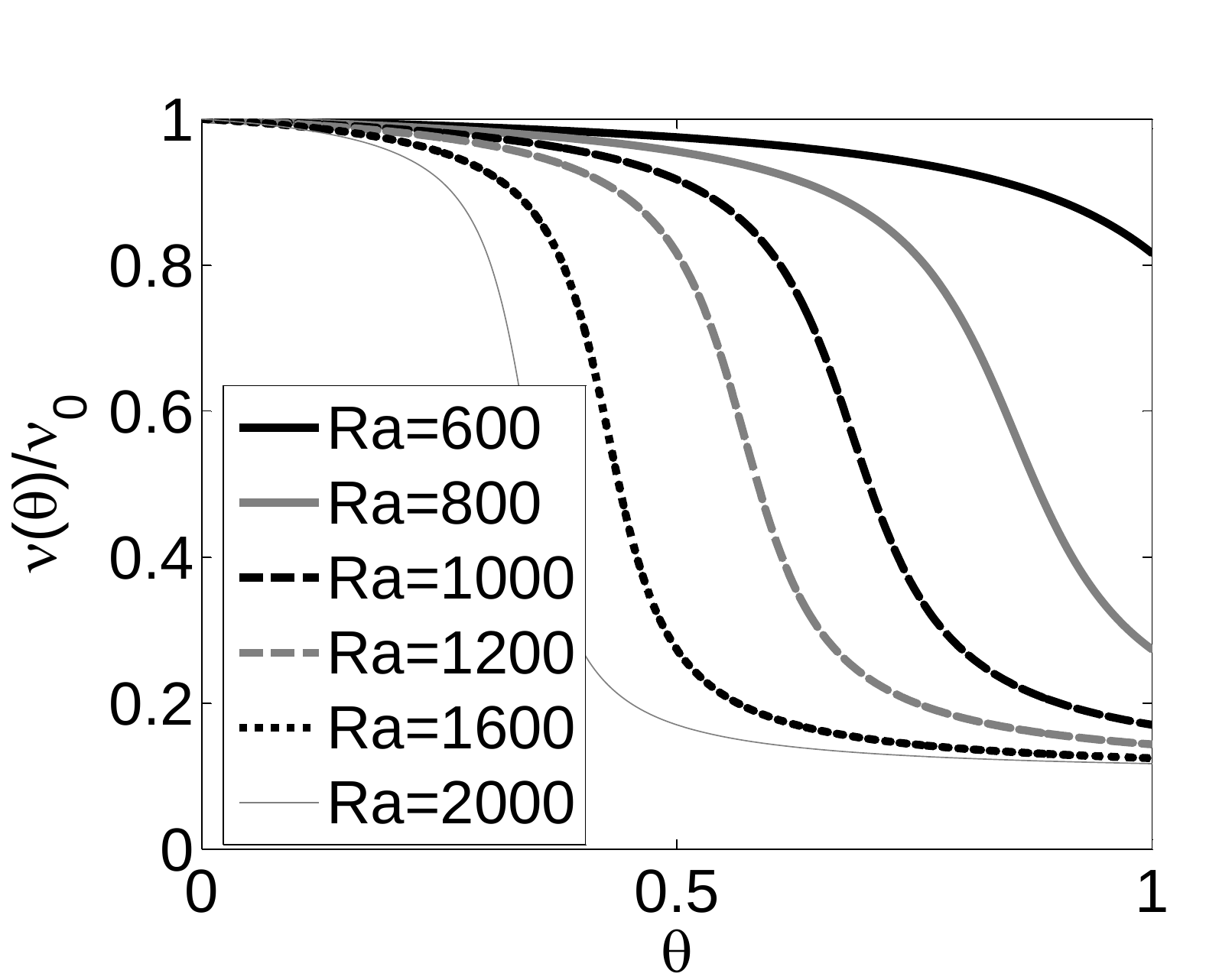}\label{Fnua}}%{a01b10nu_diferentesR.pdf}
        \subfigure[]{\includegraphics[width=4.5cm]{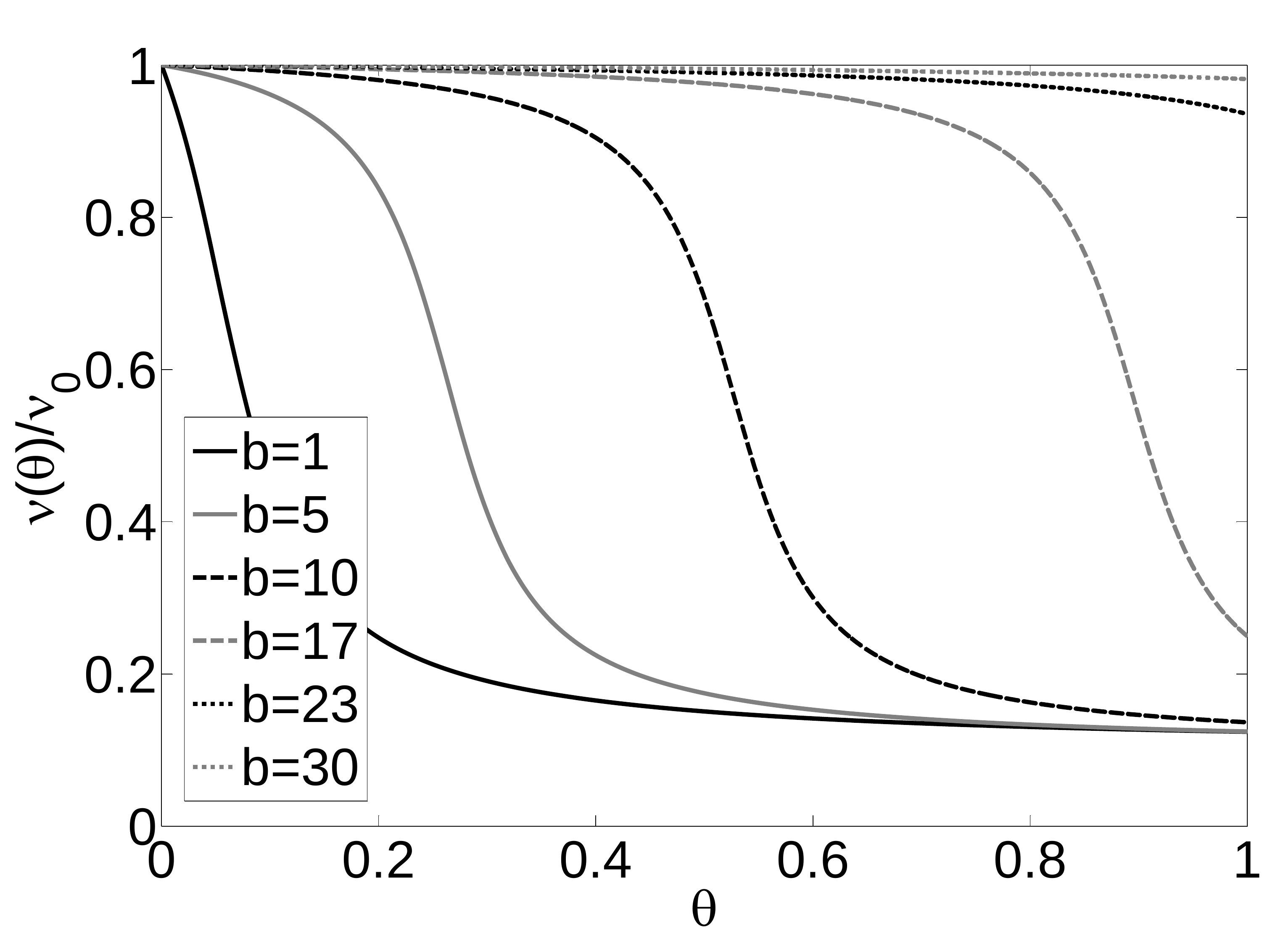}\label{Fnub}}
	   \subfigure[]{\includegraphics[width=4.5cm]{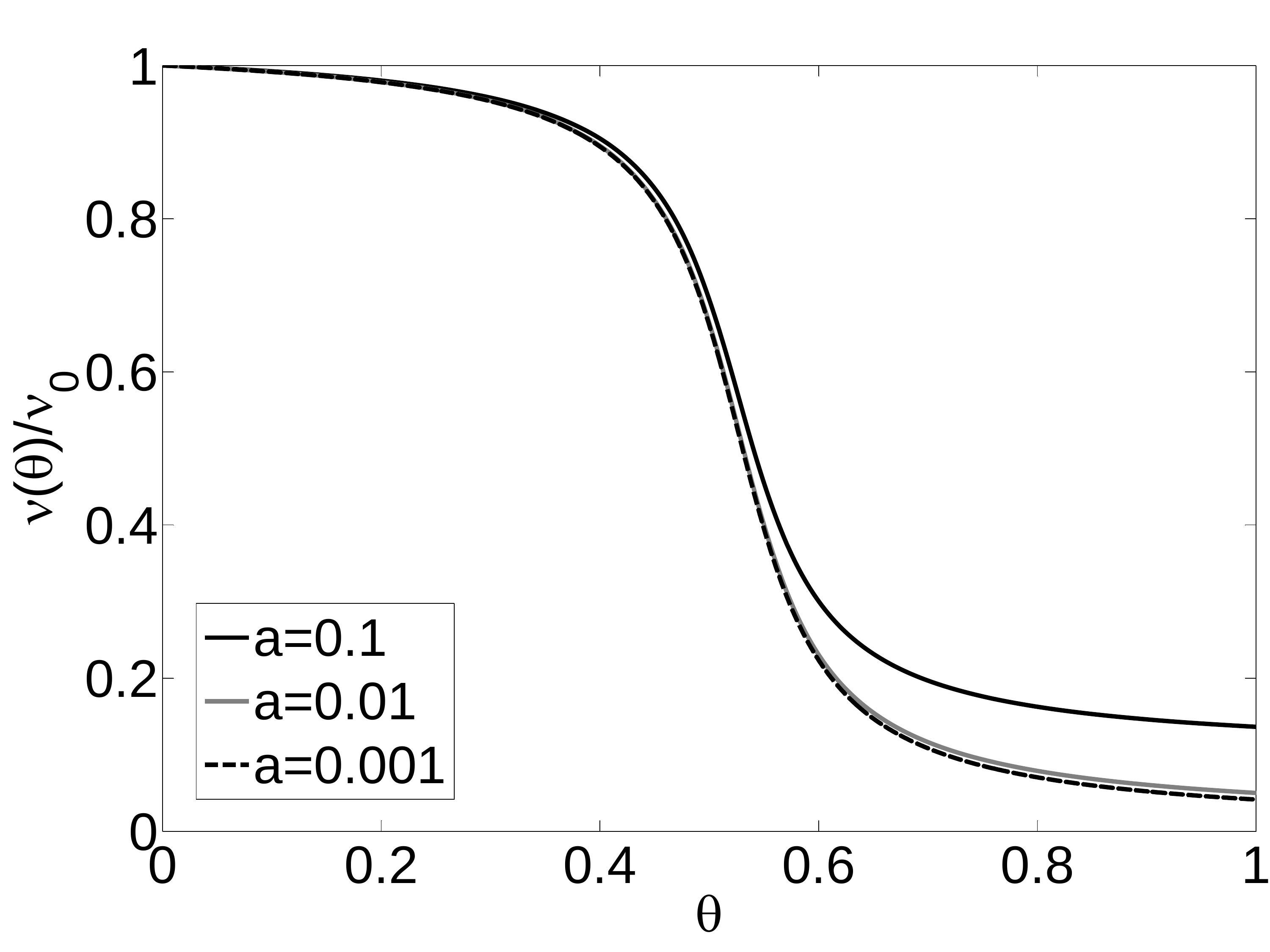}\label{Fnuc}}
	   % \label{Fnutheta}
        \caption{Representation of the arctangent viscosity law  versus the dimensionless temperature for different parameters values; a) $b=10$, $a=0.1$ and different $\Ra$ values; b)  $a=0.1$, $\Ra=1300$ and different $b$ values; c)  $b=10$, $\Ra=1300$ and different  $a$ values.}
        \label{Fnu}
   \end{figure}

\begin{figure}
    \subfigure[]{\includegraphics[width=6.5cm]{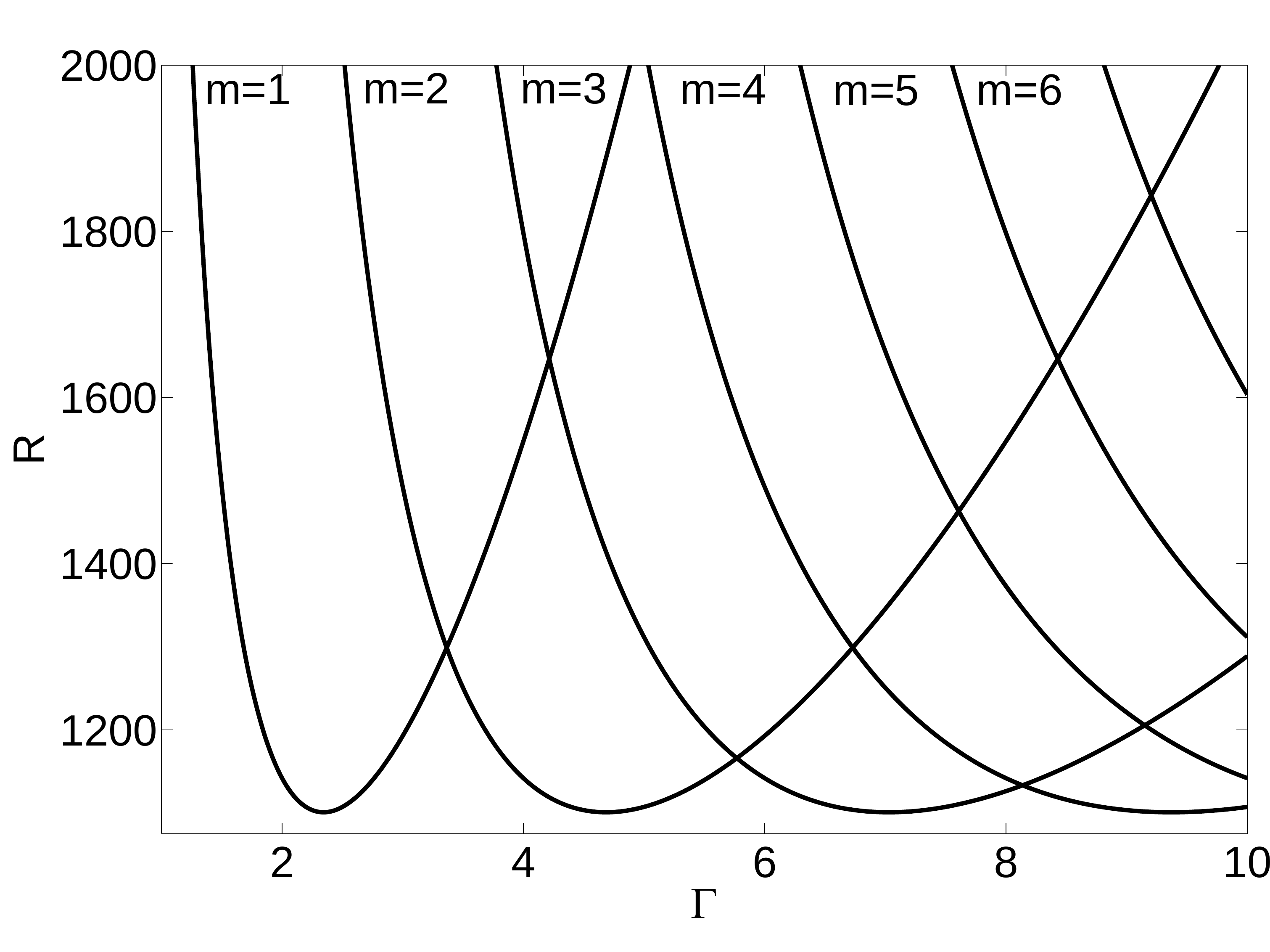}\label{CICa}}
    \subfigure[]{\includegraphics[width=6.5cm]{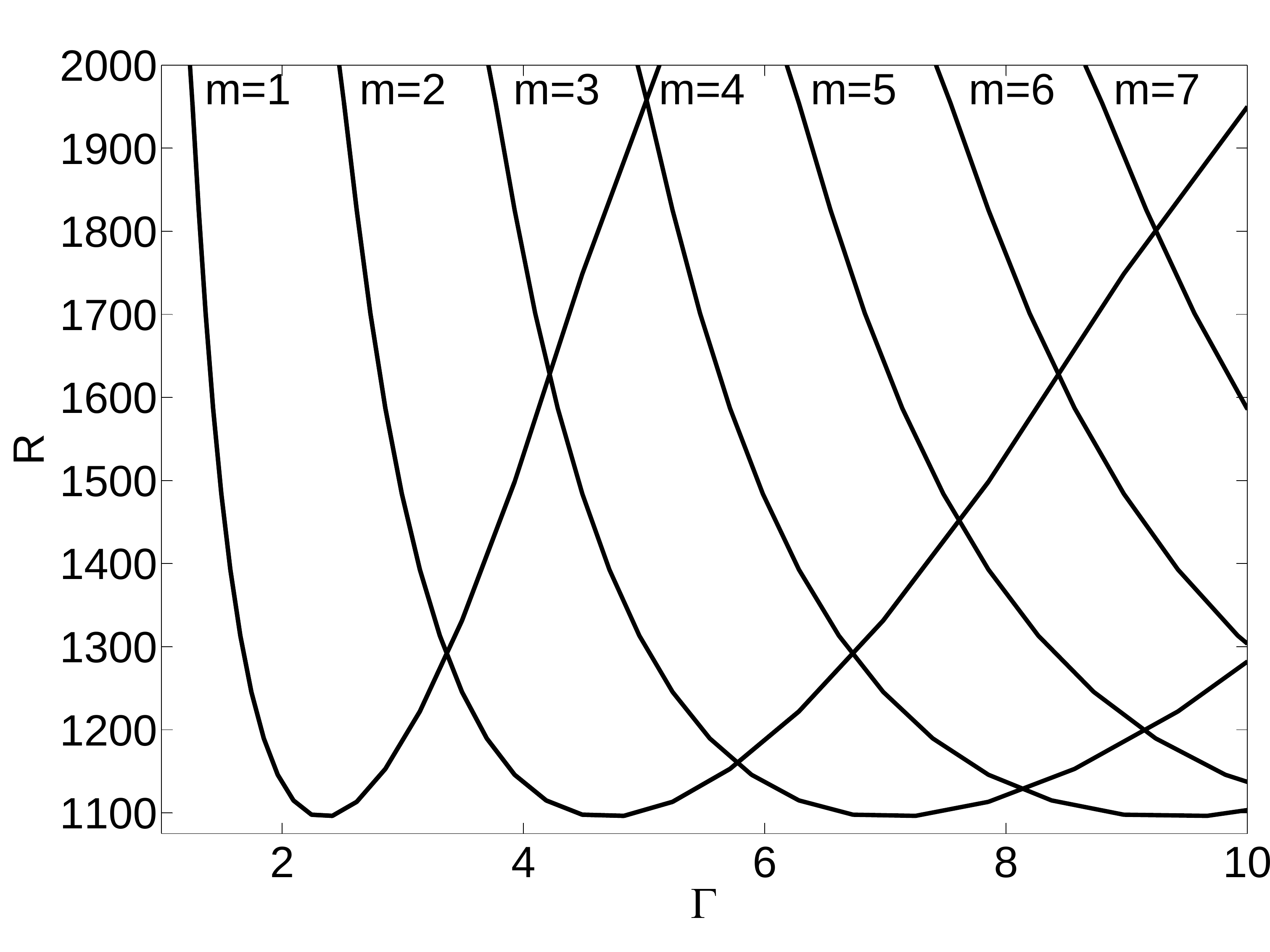}\label{CICb}}
    \caption{Critical instability curves    {of the Rayleigh number, $\Ra$, versus the aspect ratio $\Gamma$ at different wave numbers $m$. The results are } for a fluid layer a) with constant viscosity; b) with temperature dependent viscosity $\mu=0.0146$ $a=0.1$ and $b=30$.}\label{CIC}
   \end{figure}

\begin{figure}
 \centering
   \includegraphics[width=12cm]{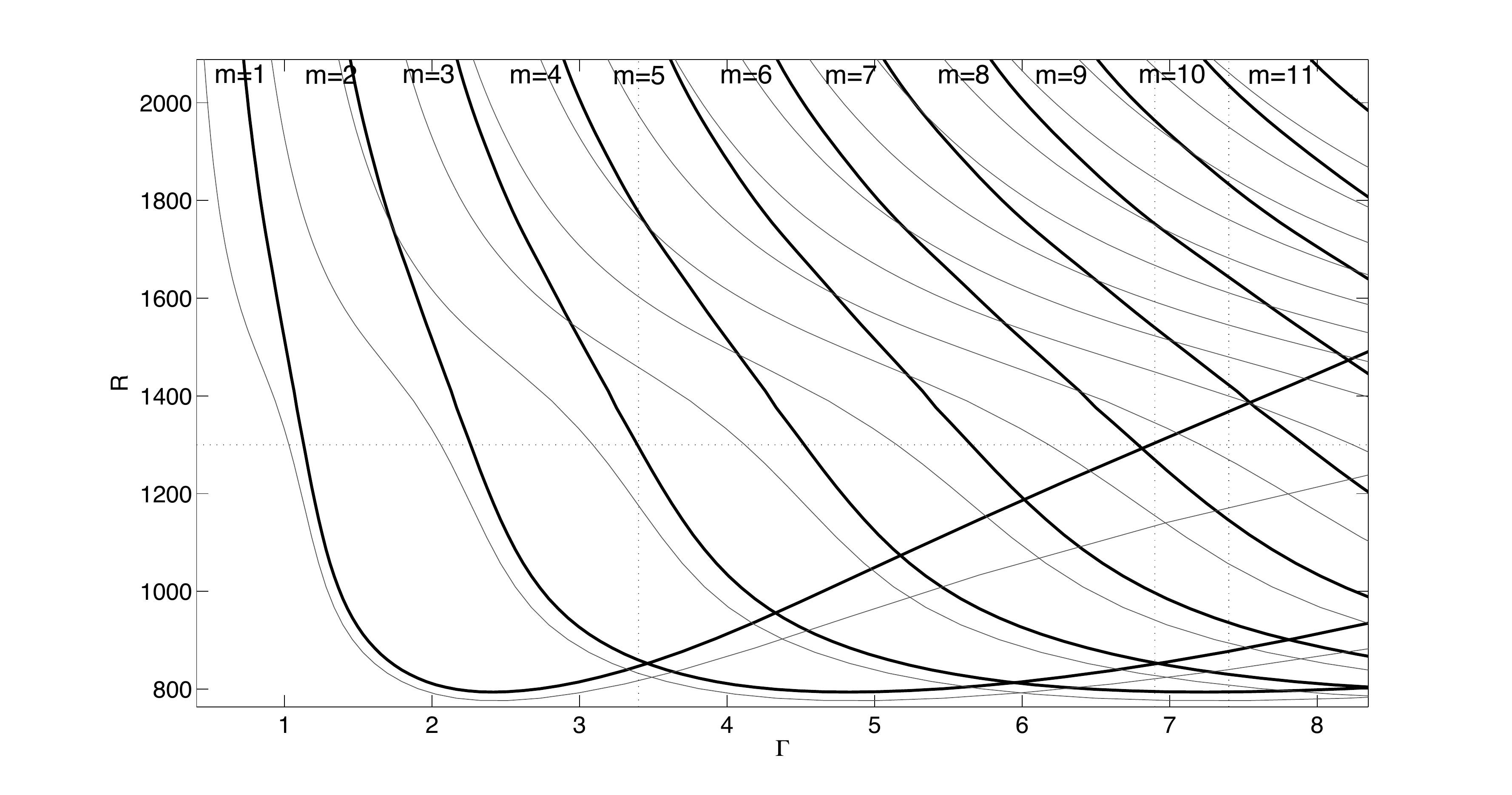}
    \caption{Critical instability curves    {of the Rayleigh number, $\Ra$, versus the aspect ratio $\Gamma$ at different wave numbers $m$. The results are } for a fluid layer with temperature dependent viscosity $\mu=0.0146$,  {$b=10$  and $a=0.1$ (thick line) or $a=0.01$ (thin line).} }\label{F2}
   \end{figure}

We consider that the viscosity  depends on temperature, and that it  changes more or less abruptly at  a certain temperature interval centered at a temperature of transition.
This is expressed with an arctangent law which reads as follows:
\begin{equation}
\label{eqarcotangd}
{\nu(T)}= A_1 \arctan(\beta \{ (T-T_1) - b \})+A_2 \end{equation}
 The parameter $\beta$ controls how abrupt the transition of the viscosity with temperature is. Very high $\beta$ values imply that the viscosity transition occurs within a very narrow temperature gap, while a finite and not too large value $\beta$ assumes that the phase change happens over a
mushy region of finite thickness \cite{ULCRT12}.  For the results reported in this article, we have fixed  $\beta=0.9$. As $\beta$ is fixed, the viscosity transition  always occurs
 in a temperature interval with constant  thickness $\Delta \theta\sim0.23$.
The temperature at which the transition occurs is controled by  $b$. The constants $A_1$ and $A_2$ are adjusted by imposing that at the reference temperature $T_1$
the viscosity law  (\ref{eqarcotangd}) must be $\nu_0$. On the other hand, in the limit $T>>T_1$, for instance $T-T_1=2500$, the viscosity is fixed to a fraction $a$ of the viscosity $\nu_0$.
These conditions supply  the system:
 \begin{eqnarray}
 {\nu_0}&=& A_1 \arctan(-\beta b )+A_2 \nonumber\\
 {\nu_0 a}&=& A_1 \arctan(\beta \{ 2500 - b \})+A_2\nonumber
 \end{eqnarray}
 which has the solution:
 \begin{align*}&A_1=  \frac{\nu_0(1-a)}{\arctan(-\beta b)- \arctan(\beta(2500-b))}, \\
&A_2= \nu_0-A_1 \arctan (-b \beta).\end{align*}
In dimensionless form, the viscosity law becomes:
 \begin{equation}\label{eqarcotang} \frac{\nu(\theta)}{\nu_0}= C_1 \arctan(\beta(\Ra\theta \mu - b))+C_2 \end{equation}
where $C_1=A_1/\nu_0$ and  $C_2=A_2/\nu_0$. In this expression, $\Ra$ is the Rayleigh number, $\theta$ is the dimensionless temperature, which takes values between 0 at the upper surface and 1 at the bottom. The parameter $\mu$, defined as $\mu=\nu_0 \kappa/(d^3\alpha g)$,     is in this study fixed  to  $\mu=0.$0146.
The parameter  $a$ is related to the inverse of the  maximum viscosity contrast on the fluid layer, although the viscosity $\nu_0 a$ may not correspond
to any element of  the fluid layer. For instance Figure  \ref{Fnua} shows  the viscosity variation with temperature  for different Rayleigh numbers at $a=0.1$ and $b=10$. It is observed that at low $\Ra$, $\Ra=600$, the viscosity is almost uniform in the fluid layer,  and it is only beyond $\Ra=1000$ that the sharp change in the viscosity is perceived.   Figure  \ref{Fnub} shows the effect of varying $b$ at $\Ra=1300$ and $a=0.1$.  If $b$ is as small as 1, the transition occurs close to $\theta=1$ and most of the layer has low viscosity, while if $b$ is very large at this $\Ra$ number most of the fluid has constant viscosity $\nu_0$.
It is   interesting to relate the viscosity law as represented in these figures with the linear stability analysis of a fluid layer with constant viscosity $\nu_0$,  as presented in Figure \ref{CIC}. In this figure, one may observe that the critical $\Ra$ number is approximately $\Ra_c\sim 1100$. On the other hand, in Figure  \ref{Fnub} one may observe  that if $b$ is large,  the viscosity near the critical Rayleigh number is almost constant across the  fluid layer. In this case, the phase transition is noticed in the fluid at large $\Ra$ numbers,  well above $\Ra=1300$, in a convection state in which vigorous plumes are already formed, as may be deduced from Figure  \ref{Fnua}. Figure  \ref{CICa} confirms that at this limit the instability threshold of the conductive state remains unchanged with respect to that obtained with constant viscosity.  On the other hand, if $b$ is small, changes in the fluid viscosity are noticed at low $\Ra$ numbers --below the critical threshold  of  a fluid with
 constant viscosity--
and in this case the instability threshold of the conductive state is affected by the phase transition. This is illustrated,
for instance, in Figures \ref{Fnua} and  \ref{CICb}. For $b=10$ and $a=0.1$, the changes in the viscosity across the fluid layer are noticed from $\Ra=800$ onwards, which is below the instability threshold obtained for constant viscosity. In this case, the instability thresholds for the conductive solution are as those displayed in Figure  \ref{F2}, and thus the phase transition is perceived from the beginning by weakly convective states.

\begin{figure*}
  \centering
   \includegraphics[width=8.5cm]{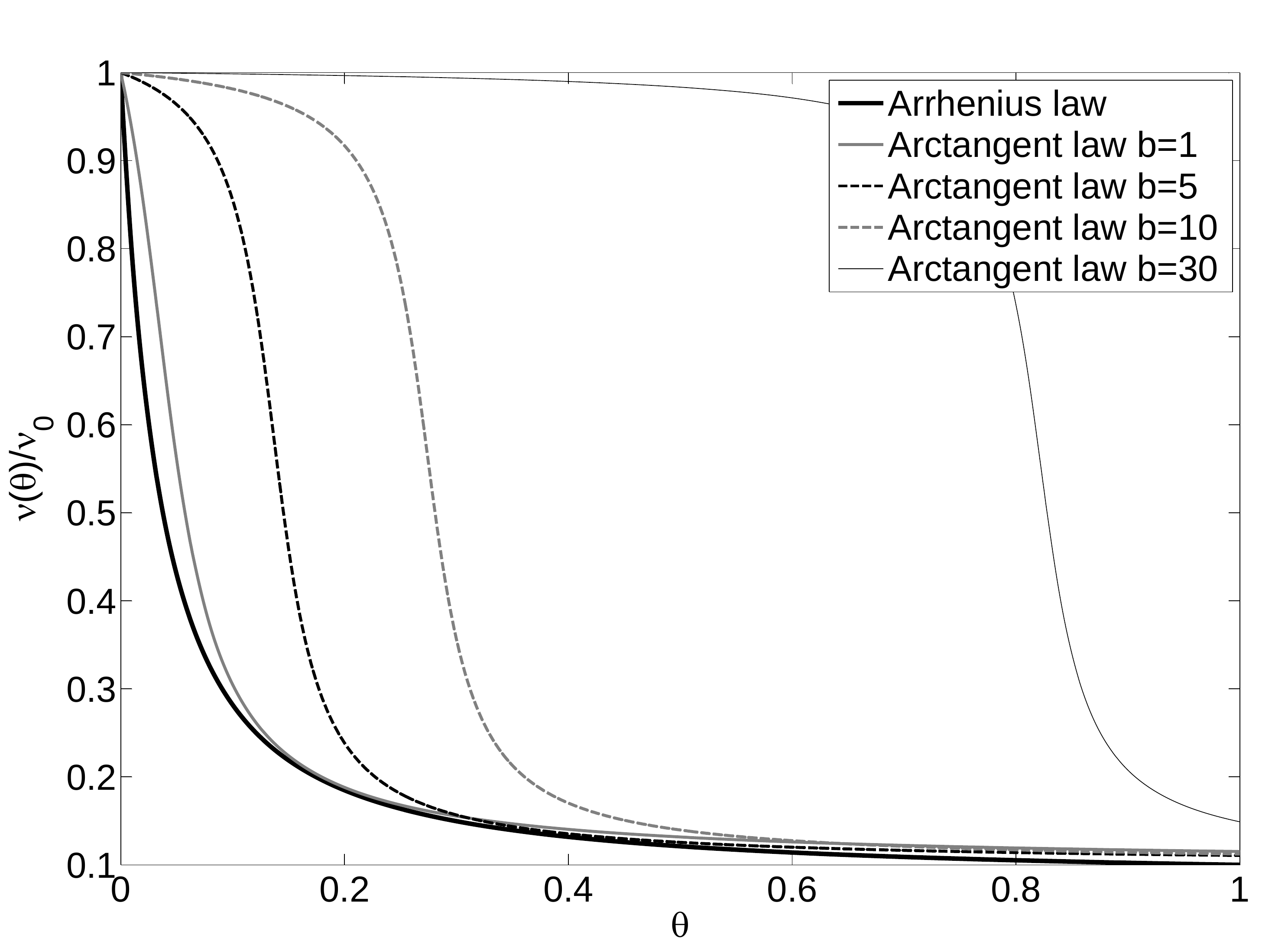}
    \caption{The law of the viscosity dependent on the temperature used in \cite{KK97} with viscosity contrast of factor 10 against the arctangent law \eqref{eqarcotang} with parameters $b=1,5,10,30$, $\Ra=2500$ and $a=0.1$ }\label{FArrhenius}
   \end{figure*}

We now discuss the relation between  the arctangent law and an  Arrhenius type law frequently used in the literature to model mantle convection problems. This viscosity law
  is expressed  according to \cite{KK97, Dav01} as:
    \begin{equation}
 \nu(\theta)=\nu_0 \exp\left[ \frac{E^*}{\overline{R}\Delta \theta}\left(\frac{1}{\theta+t_1}-\frac{1}{1+t_1} \right)\right]\label{eqarr}
 \end{equation}
where $E^*$ is the activation energy, $\overline{R}$ is the universal gas constant,  $\Delta \theta$ is the temperature drop across the fluid layer and $t_1$ is the surface temperature divided by the temperature drop across the layer. Figure \ref{FArrhenius}  represents the viscosity (\ref{eqarr}) versus the dimensionless temperature for
$\frac{E^*}{\overline{R}\Delta \theta}=0.25328$ and $t_1=0.1$ as considered by \cite{KK97}. Additionally, several arctangent laws with different $b$ values are displayed. In this representation, one may observe the great similitude between the Arrhenius law and the arctangent law for $b=1$. At larger $b$ values, the decaying rate between viscosities is still similar to an Arrhenius law; however,  temperature intervals exist with approximately constant viscosities $\nu_0$ and $\nu_0a$.

{One of the effects of the viscosity contrasts in the fluid motion is that if they are very large, as  achieved for instance with the exponential or the Arrhenius law,  they lead to a stagnant lid convection regime \cite{MS95, SM96, SM97}, in which there exists a non mobile cap where heat is dissipated mainly by conduction over a convecting flow. In \cite{ULCRT12,CM13} a similar stagnant regime is obtained for a  viscosity law similar to the one presented in this section. In our setting, we have considered a free slip boundary at the top boundary, thus quiescence is not imposed. This condition enables us to consider spontaneous transitions from stagnant to  non stagnant regimes.}
%As we have imposed that the viscosity in the upper boundary must be $\nu_0$ always occur a transition in the fluid interior, no matter how large $R$ is.

 \section{Numerical methods}\label{SSolutions}

Analysis of the solutions  to the problem described by equations (\ref{eqproblem1})-(\ref{eqproblem3}) and boundary conditions (\ref{eqbc})  is assisted by
 time dependent numerical simulations and  bifurcation techniques such as branch continuation. As highlighted by  \cite{CuMa11,PMH09}, the  combination of both techniques  provide a thorough insight into the solutions observed in the system.  A full discussion on  the  spectral numerical schemes used is given  in \cite{CuMa11}.
 For completeness, we now summarize
 the essential elements of the numerical approach.

\subsection{Stationary solutions and their stability}

 The simplest stationary solution to the problem described by equations (\ref{eqproblem1})-(\ref{eqproblem3}) with boundary conditions (\ref{eqbc}) is the
  conductive solution which satisfies $\mathbf{u}_c=0$ and $\theta_c=-z+1$.  This solution is stable only  for a range of  vertical temperature gradients which are represented by
small enough Rayleigh numbers. Beyond the
critical threshold $\Ra_c$,  a convective motion settles in and new structures are observed which
may be either time dependent or stationary. In the latter case, the stationary equations, obtained by canceling the time derivatives in the system  (\ref{eqproblem1})-(\ref{eqproblem3}) are satisfied
by the bifurcating solutions.  {At the instability threshold of the conductive state, the growing solutions are periodic and correspond to sine or cosine eigenfunctions with wave number $m$. Figures \ref{CIC} and \ref{F2} display the critical instability curves for different $m$ values as a function of the aspect ratio.}
The new solutions depend on the external
physical parameters, and new critical thresholds exist at which  stability is lost, thereby giving rise to new bifurcated structures.
These solutions are numerically obtained  by using  an iterative Newton-Raphson method. This method starts  with an  approximate solution at step $s=0$,  to which
 is added a small correction in tilde:
 \begin{equation}
 ({\bf u}^s+ \tilde{{\bf u}},  \theta^s + \tilde{\theta},  P^s + \tilde{P}). \label{pertu}
 \end{equation}
These expressions are introduced into the system (\ref{eqproblem1})-(\ref{eqproblem3}), and after canceling the nonlinear terms in tilde, the following equations are obtained:
\begin{align}
0=&\nabla \cdot \tilde{\mathbf{u}}+\nabla \cdot \mathbf{u}^s,\label{eqtem13}\\
0=&-\partial_x \tilde{P} -\partial_x P^s +\frac{1}{\nu_0} [L_{11}(\theta^s, u_x^s,u_z^s)+L_{12}(\theta^s)\tilde{u}_x \nonumber\\
&+L_{13}(\theta^s)\tilde{u}_z +L_{14}(\theta^{s},u_x^s, u_z^s)\tilde{\theta}],\label{eqtem23}\\
0=&-\partial_z \tilde{P}-\partial_z P^s+\frac{1}{\nu_0} [L_{21}(\theta^s, u_x^s,u_z^s)+L_{22}(\theta^s)\tilde{u}_x \nonumber\\
&+L_{23}(\theta^s)\tilde{u}_z +(L_{24}(\theta^{s},u_x^s, u_z^s)+\Ra)\tilde{\theta}],\label{eqtem33}\\
0=&\tilde{\mathbf{u}}\cdot \nabla \theta^{s}+\mathbf{u}^{s}\cdot \nabla \tilde{\theta}+\mathbf{u}^{s}\cdot \nabla \theta^{s}-\Delta \tilde{\theta} -\Delta {\theta^s}.\label{eqtem43}
\end{align}
Here, $L_{ij}$ ($i=1,2$, $j=1,2,3,4$) are linear operators with non-constant coefficients, which are defined as follows:
\begin{align}
L_{11}(\theta, u_x, u_z)=&2 \partial_{\theta} \nu(\theta) \partial_x \theta \partial_x u_x + \nu(\theta)\Delta u_x+ \partial_{\theta}\nu(\theta) \partial_z \theta (\partial_x u_z + \partial_z u_x),\label{eqL11}\\
L_{12}(\theta)=& 2\partial_{\theta} \nu(\theta)\partial_x \theta \partial_x + \nu(\theta)\Delta + \partial_{\theta} \nu(\theta)\partial_z \theta \partial_x ,\label{eqL12}\\
L_{13}(\theta)=& \partial_{\theta}\nu(\theta)\partial_z \theta\partial_x,\label{eqL13}\\
L_{14}(\theta,u_x,u_z)=& 2\partial_{\theta} \nu(\theta)\partial_x u_x\partial_x +2\partial^2_{\theta \theta}\nu(\theta) \partial_x \theta \partial_x u_x +\partial_{\theta}\nu(\theta) \Delta u_x \nonumber\\&+  (\partial_x u_z + \partial_z u_x)(\partial_{\theta} \nu(\theta)\partial_z +\partial^2_{\theta\theta}\nu(\theta)\partial_z \theta ),\label{eqL14}\\
L_{21}(\theta, u_x, u_z)=&2 \partial_{\theta} \nu(\theta) \partial_z \theta \partial_z u_z + \nu(\theta)\Delta u_z +\partial_{\theta}\nu(\theta) \partial_x \theta (\partial_z u_x + \partial_x u_z),\label{eqL21}\\
L_{22}(\theta)=&\partial_{\theta}\nu(\theta)\partial_x \theta\partial_z ,\label{eqL22}\\
L_{23}(\theta,u_x,u_z)= & 2\partial_{\theta} \nu(\theta)\partial_z \theta \partial_z + \nu(\theta)\Delta + \partial_{\theta} \nu(\theta)\partial_x \theta \partial_z ,\label{eqL23}\\
L_{24}(\theta,u_x,u_z)=& 2\partial_{\theta} \nu(\theta)\partial_z u_z\partial_z  +2\partial_{\theta \theta}\nu(\theta) \partial_z \theta \partial_z u_z +\partial_{\theta}\nu(\theta)\Delta  u_z  \nonumber\\
&+  (\partial_z u_x + \partial_x u_z)(\partial_{\theta} \nu(\theta)\partial_x +\partial_{\theta\theta}\nu(\theta)\partial_x \theta).\label{eqL24}
\end{align}
 The unknown fields
$\tilde{{\bf u}}, \,\, \tilde{P}, \,\, \tilde{\theta}$ are found by solving the linear system with the boundary conditions:
\begin{align}\label{eqbcp}
\tilde{\theta}=0, \ \tilde{\mathbf{u}}=\vec{0},\text{ on } z=0 \text{ and } \
\tilde{\theta}=\partial_z \tilde{u_x}=\tilde{u_z}= 0,\text{ on } z=1.
\end{align}
Then the new approximate solution $s+1$ is  set to
\begin{equation} {\bf u}^{s+1}= {\bf u}^s+ \tilde{{\bf u}}, \;  \theta^{s+1}= \theta^s+ \tilde{\theta}, \;  P^{s+1}= P^s+ \tilde{P}.\nonumber\end{equation}
The whole procedure is repeated for $s+1$ until a convergence criterion  is fulfilled. In particular,
we  consider
that the $l^{2}$ norm of the computed perturbation
should be less than $10^{-9}$.

The study of the stability  of the stationary solutions under consideration is  addressed  %as for the conductive solution
 by means of a linear stability analysis. Now perturbations are added to a general
stationary solution, labeled with superindex $b$:
\begin{eqnarray}
\mathbf {u}(x,z,t)&=&\mathbf {u}^b(x,z)+\tilde{\mathbf{u}} (x,z){\rm e}^{\lambda t}
\label{pert1},\\
\theta(x,z,t)&=&\theta^b(x,z)+\tilde{\theta}(x,z){\rm e}^{\lambda t},\label{pert2}\\
P(x,z,t)&=&P^b(x,z)+\tilde{P}(x,z){\rm e}^{\lambda t}.\label{pert3}
\end{eqnarray}
The sign in the real part of the eigenvalue $\lambda$ determines the stability of
the solution:  if it is  negative, the perturbation decays and the stationary solution
is stable, while if it is positive the perturbation grows over time and the conductive solution
is unstable. The linearized equations are:
\begin{align}
0=&\nabla \cdot \tilde{\mathbf{u}} \label{eqtem13_2}\\
0=&-\partial_x \tilde{P}  +\frac{1}{\nu_0} [L_{12}(\theta^b)\tilde{u}_x +L_{13}(\theta^b)\tilde{u}_z +L_{14}(\theta^{b},u_x^b, u_z^b)\tilde{\theta}]\label{eqtem23_2}\\
0=&-\partial_z \tilde{P}+\frac{1}{\nu_0} [L_{22}(\theta^b)\tilde{u}_x+L_{23}(\theta^b)\tilde{u}_z +(L_{24}(\theta^{b},u_x^b, u_z^b)+\Ra)\tilde{\theta}]\label{eqtem33_2}\\
0=&\tilde{\mathbf{u}}\cdot \nabla \theta^{b}+\mathbf{u}^{b}\cdot \nabla \tilde{\theta} +\mathbf{u}^{b}\cdot \nabla {\theta^b}-\Delta \tilde{\theta}+\lambda \tilde{\theta},\label{eqtem43_2}
\end{align}
where the operators $L_{ij}$ are the same as those defined in {equations} (\ref{eqL11})-(\ref{eqL24}).  Equations (\ref{eqtem13_2})-(\ref{eqtem43_2}) jointly with its boundary conditions (identical to (\ref{eqbcp})) define a generalized eigenvalue problem.

The unknown fields $Y$ of the stationary (\ref{eqtem13})-(\ref{eqtem43}) and eigenvalue problems (\ref{eqtem13_2})-(\ref{eqtem43_2})  are approached by means of a spectral method according to the expansion:
\begin{align}\label{eqexpansion3}
Y(x,z)=&\sum_{l=1}^{\lceil L/2\rceil}\sum_{m=0}^{M-1} b^Y_{lm}T_m(z)\cos((l-1)x) + \sum_{l=2}^{\lceil L/2\rceil}\sum_{m=0}^{M-1} c^Y_{lm}T_m(z)\sin((l-1)x).
\end{align}
In this  notation,  $\lceil \cdot \rceil$ represents the nearest integer towards infinity. Here $L$ is an odd number as justified in \cite{CuMa11}. $4\times L\times M$ unknown coefficients exist  which are determined by a collocation method in which equations and boundary conditions
are imposed at the collocation points $(x_j,z_i),$
\begin{align*}
\text{Uniform grid:}\quad &x_j= (j-1)\frac{2\pi}{L}, &j=1,\ldots, L;&\\
\text{Gauss--Lobatto:} \quad &z_i= \cos\left( \left( \frac{i-1}{M-1}-1\right)\pi\right), & i=1,\ldots,M;&
\end{align*}
according to the rules detailed in \cite{CuMa11}. Expansion orders  $L$ and $M$ are taken to ensure accuracy on the results: details on their values are provided in the Results section.

\subsection{Time dependent schemes}
Together with boundary conditions \eqref{eqbc}, the governing equations \eqref{eqproblem1}--\eqref{eqproblem3} define a time-dependent problem for which we propose a temporal scheme based on a spectral  spatial discretization   analogous to that proposed in the previous section.
As before, expansion orders  $L$ and $M$ are such that they ensure accuracy on the results and details on their values are given in the following section.
To integrate in time, we use a  third order multistep scheme. In particular,  we use a backward differentiation formula (BDF),  adapted for use with a variable time step.
The variable time step scheme controls the step size according to an estimated error $E$ for the
 fields.  The error estimation  $E$ is based on the difference between the solution obtained with a third  and a second order scheme.
 The result of an integration at time $n+1$ is accepted if $E$ is below a certain tolerance. Details on the step adjustment are found in \cite{CuMa11}.

 BDFs are a particular case of multistep formulas which are  {\it implicit},  thus  the BDF scheme   implies solving at each time step the problem (see \cite{HNW09}):
 \begin{align}
 &0=\nabla\cdot \mathbf{u}^{n+1} \label{eqtem1}\\
 &0=\Ra\theta^{n+1}\vec{e_3}-\nabla P^{n+1}+\text{div} \left(\frac{\nu(\theta^{n+1})}{\nu_0}(\nabla \mathbf{u}^{n+1}+(\nabla \mathbf{u}^{n+1})^T)\right)\label{eqtem2}\\
 &\partial_t \theta^{n+1}=-\mathbf{u}^{n+1}\cdot \nabla \theta^{n+1}+\Delta \theta^{n+1},\label{eqtem3}
 \end{align}
where  $\partial_t \theta^{n+1}$ is replaced by a backward differentiation formula.

In \cite{CuMa11}, it has been proved that instead of solving
the fully implicit scheme  (\ref{eqtem1})-(\ref{eqtem3}), a semi-implicit scheme can produce results with a similar  accuracy and fewer CPU time requirements.
The semi-implicit scheme  approaches the nonlinear terms in {equations} (\ref{eqtem1})-(\ref{eqtem3}) by  assuming that the solution
at time $n+1$ is a small perturbation $\tilde{Z}$ of the solution at time $n$; thus,  ${\bf z}^{n+1}={\bf z}^{n}+\tilde{Z}$.
Once linear equations for $\tilde{Z}$ are derived, the equations are rewritten by replacing  $\tilde{Z}={\bf z}^{n+1}-{\bf z}^{n}$. The solution is obtained  at each step by solving the resulting linear equation for variables in time $n+1$.

 \section{Results}\label{Sresults}

 \subsection{Exploration of stationary solutions in the parameter space}

 In this section we explore how stationary solutions obtained at a low aspect ratio $\Gamma=3.4$ for the system (\ref{eqproblem1})-(\ref{eqproblem3})
 depend on the parameters $a$ and $b$ of the viscosity law (\ref{eqarcotang}).  We examine the shape and structure of the plumes in a range of Rayleigh numbers
  from  $\Ra=2500$ to $\Ra=3500$.

\begin{figure*}
   \subfigure[ $b=30$, $\Ra=2500$, $a=0.1$]{\includegraphics[width=4.5cm]{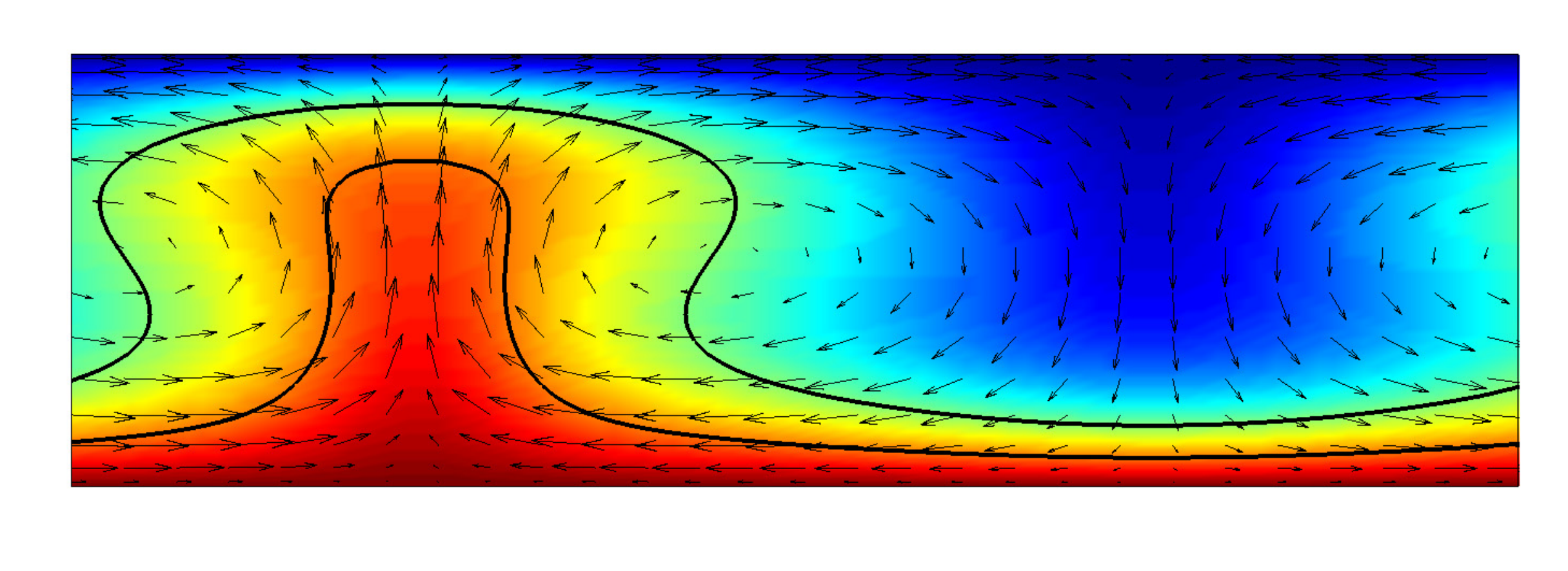}\label{b30a}}
   \subfigure[ $b=30$, $\Ra=3500$, $a=0.1$]{\includegraphics[width=4.5cm]{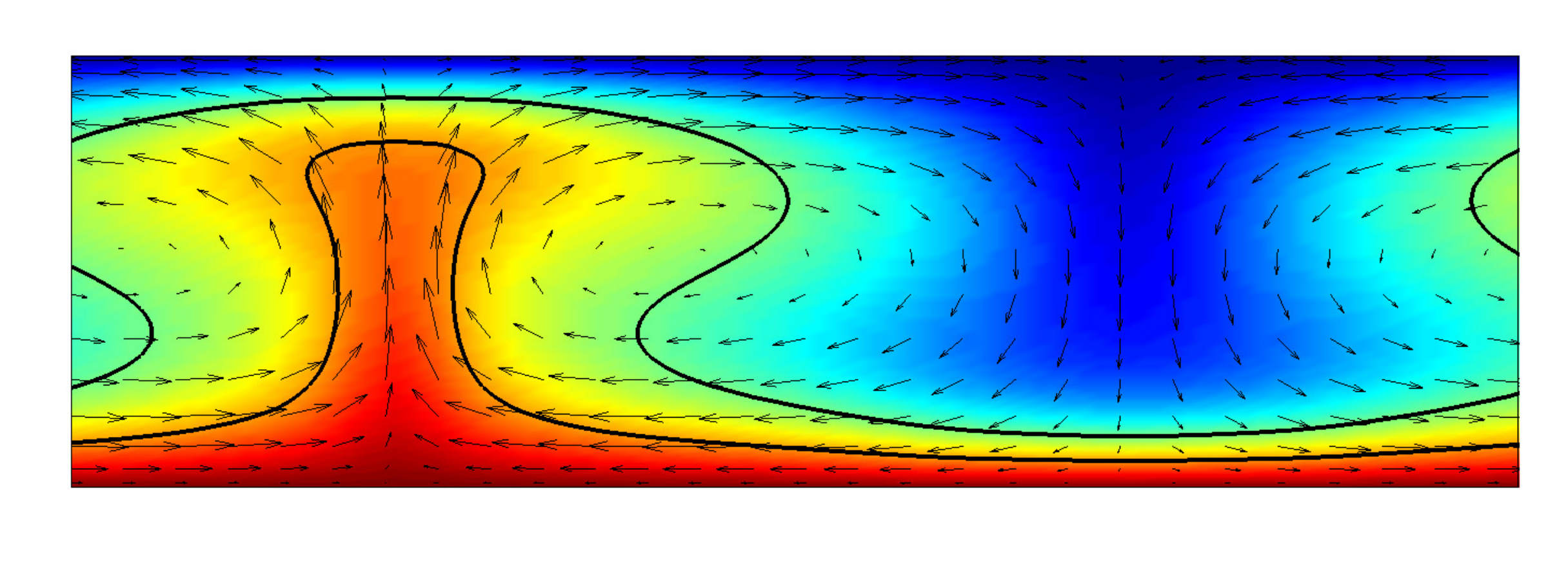}\label{b30b}}
   \subfigure[ $b=30$, $\Ra=3500$, $a=0.001$]{\includegraphics[width=4.5cm]{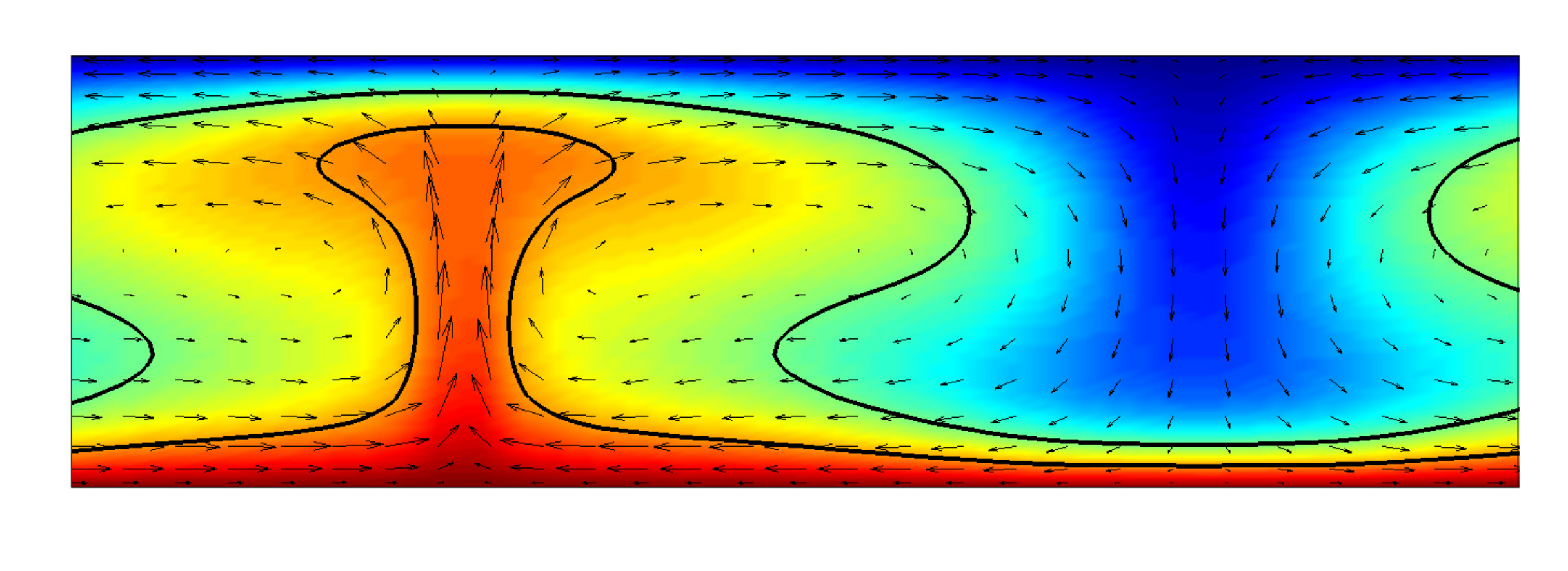}\label{b30c}}\\
   \subfigure[ $b=10$, $\Ra=2500$, $a=0.1$]{\includegraphics[width=4.5cm]{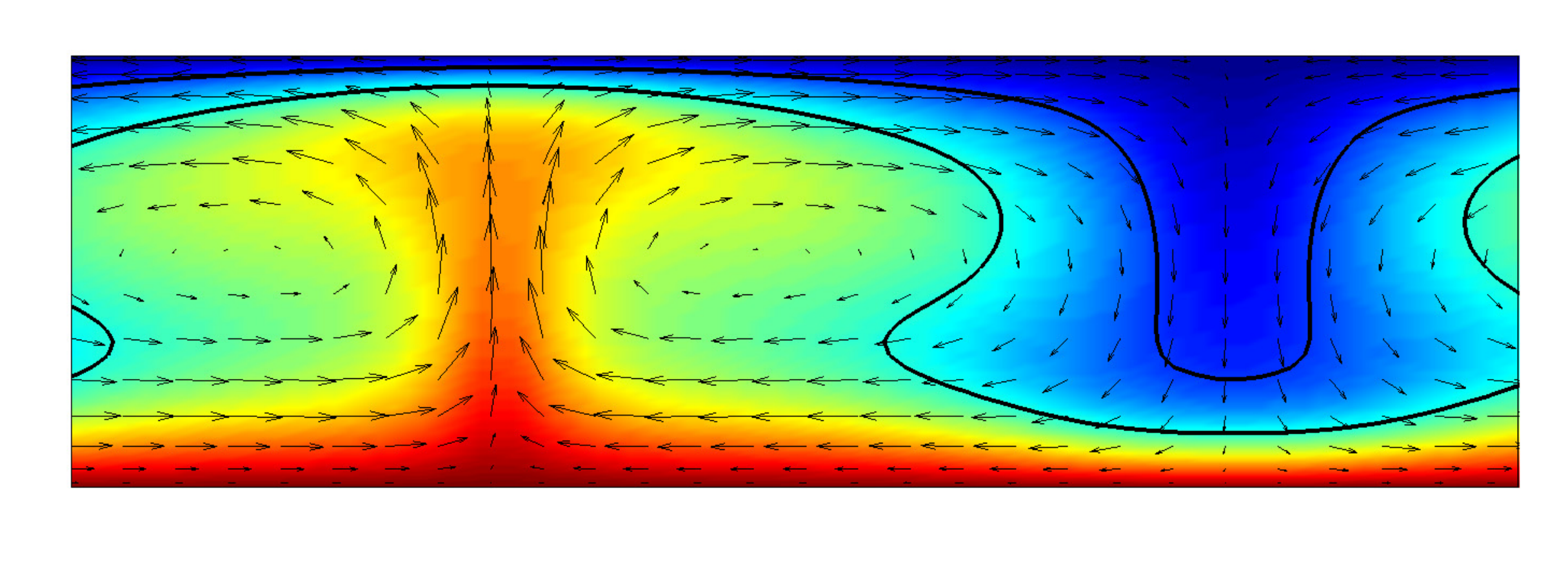}\label{b10a}}
   \subfigure[ $b=10$, $\Ra=3500$, $a=0.1$]{\includegraphics[width=4.5cm]{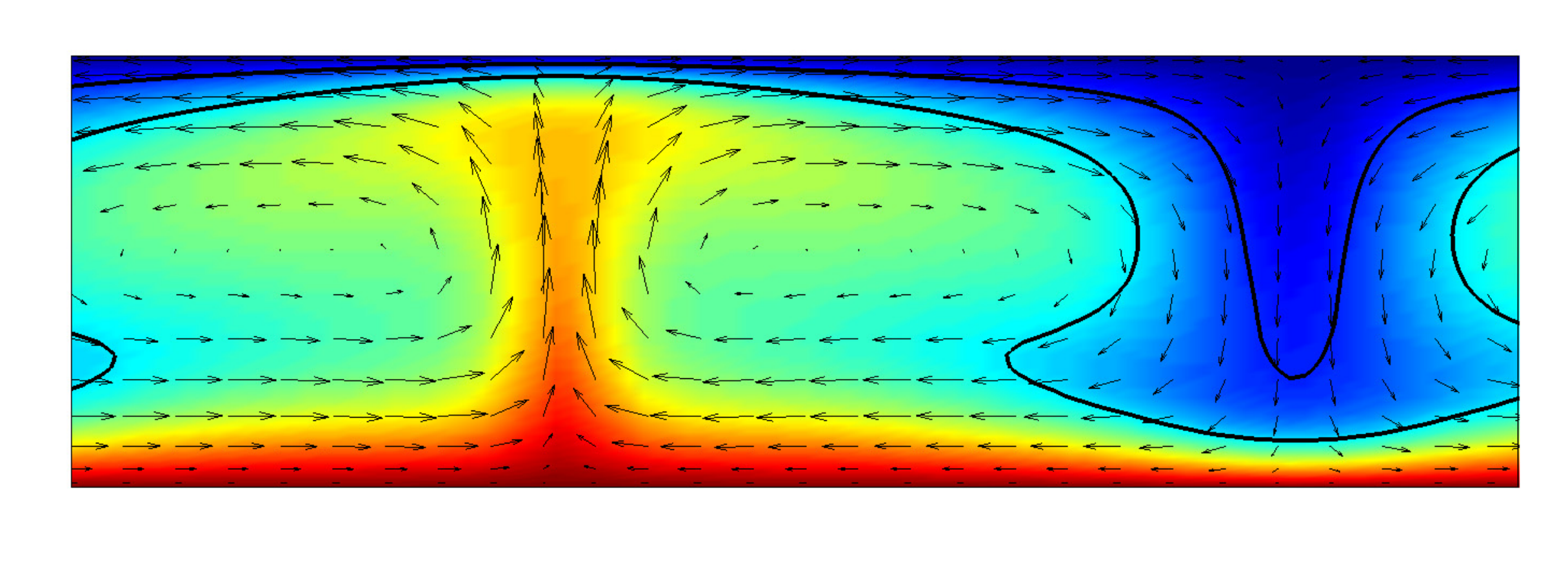}\label{b10b}}
   \subfigure[ $b=10$, $\Ra=2500$, $a=0.001$]{\includegraphics[width=4.5cm]{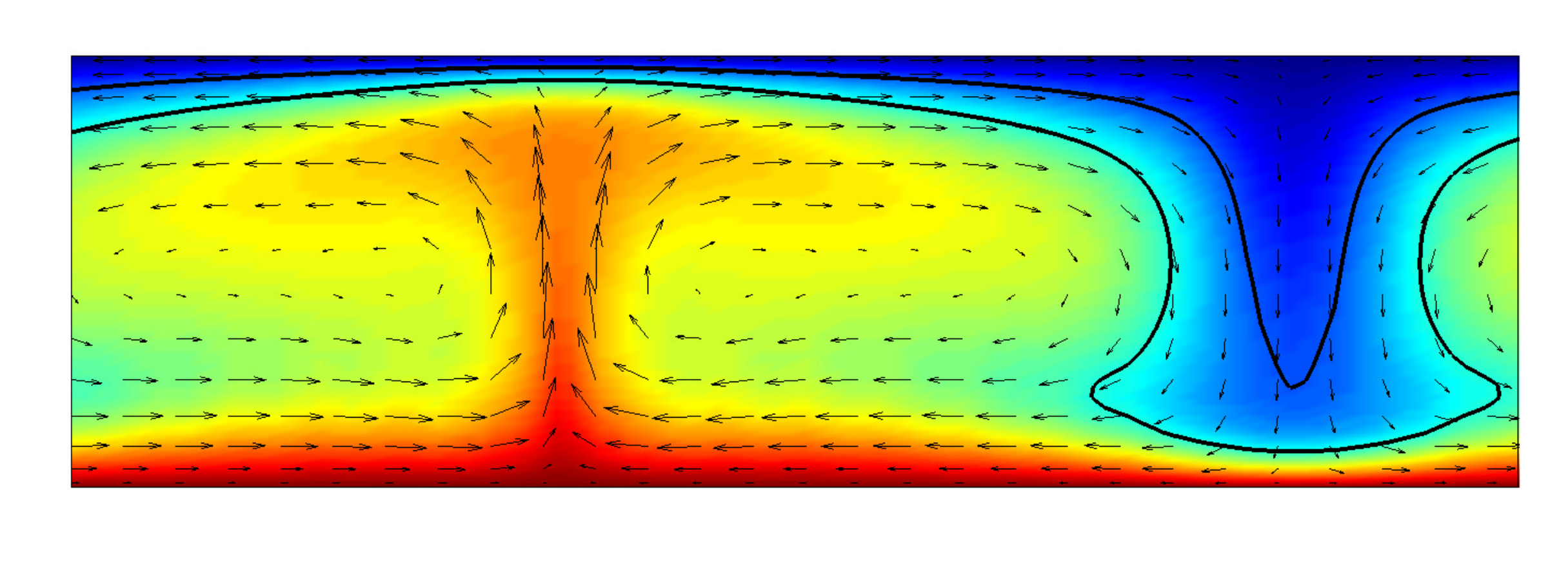}\label{b10c}}\\
   \subfigure[ $b=17$, $\Ra=2500$, $a=0.1$]{\includegraphics[width=4.5cm]{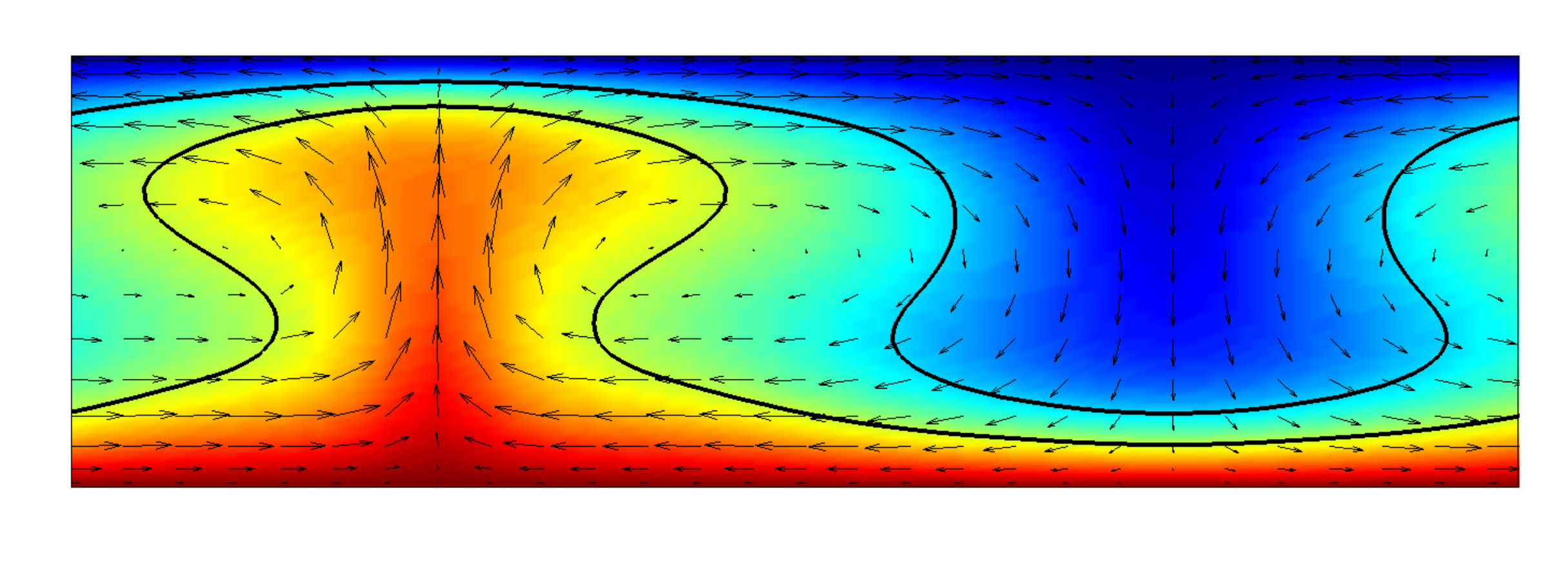}\label{b17a}}
   \subfigure[ $b=17$, $\Ra=2500$, $a=0.001$]{\includegraphics[width=4.5cm]{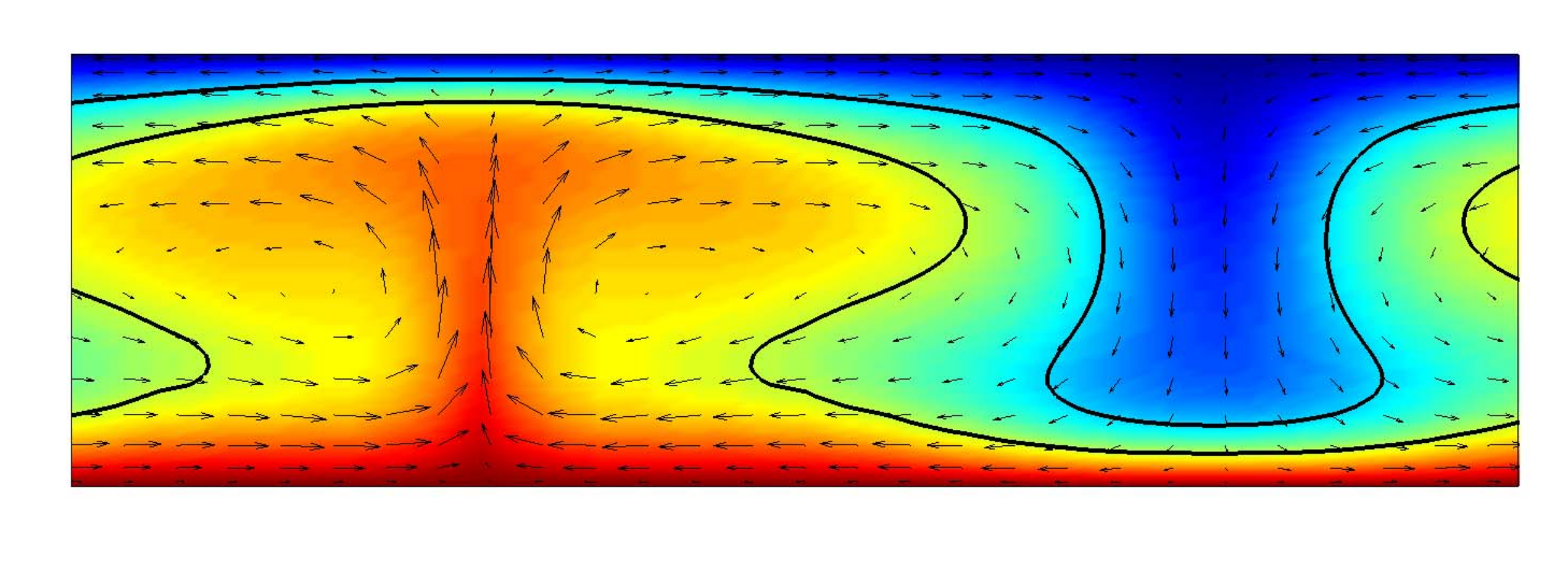}\label{b17b}}
   \subfigure[ $b=17$, $\Ra=3000$, $a=0.001$]{\includegraphics[width=4.5cm]{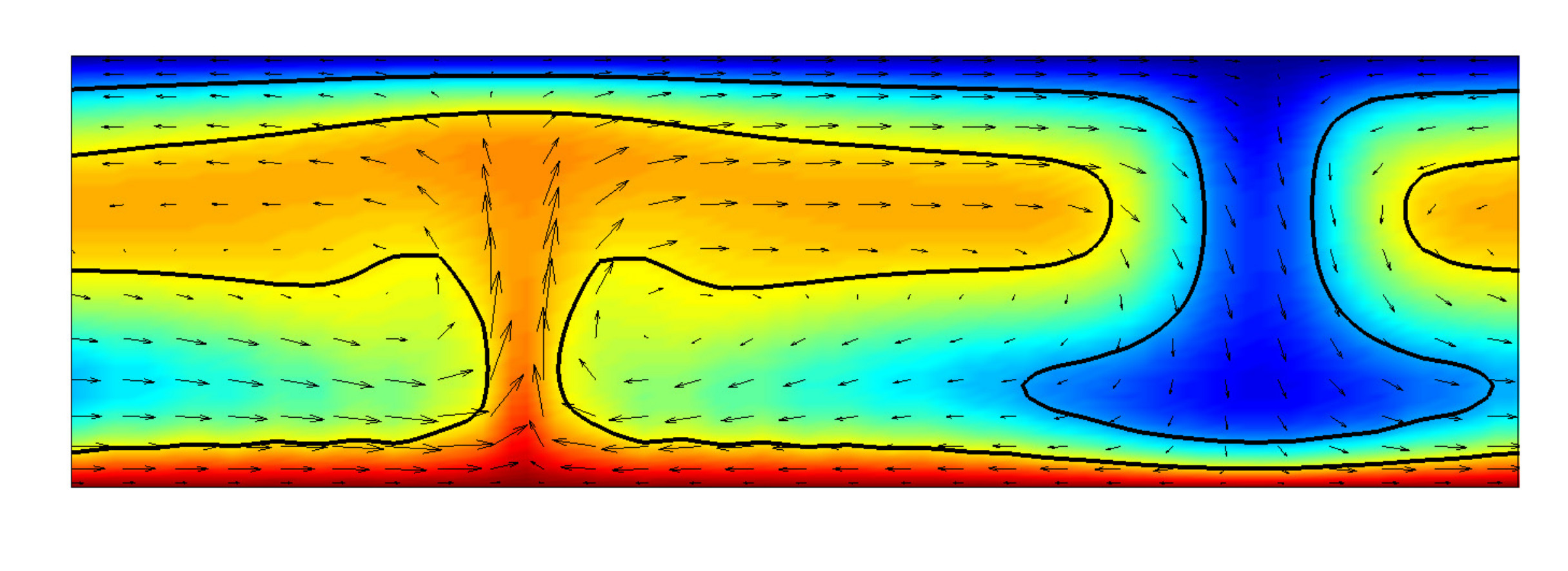}\label{b17c}}\\
   \caption{(Color online). Plumes obtained for several values of the viscosity parameter $b$. {The arrows  indicate the velocity field, while the  contour colors represent the temperature ranging from hot  (bottom plate) to cold (upper plate). The two black contour lines indicate the temperatures between which viscosity decays most rapidly. }}
   \end{figure*}

 We first consider that  the parameter $b$ is large: for instance, as large as 30. In this case, Figure \ref{Fnub}  confirms that at the instability threshold  the viscosity across the  fluid layer  is almost constant and equal to $\nu_0$,
  no matter what the value of $a$ may be. Thus, the viscosity transition becomes evident in the fluid once convection  has settled in at $\Ra$ numbers well above the instability threshold.    Figure \ref{b30a} shows the plume pattern observed at $\Ra=2500$ for $a=0.1$; although values $a=0.01$ and $a=0.001$ are not displayed, they provide a very similar output.
  The plume is  spout-shaped, with the tail of the plume  nearly as large as the head.
   In the pattern, the two black contour lines mark temperatures between which the viscosity  decays  most rapidly. These correspond to the transition region in which
 the gradient of the viscosity  law (\ref{eqarcotang})  is large. Thus one of the contours, the coldest one, fits  the temperature $\theta_1$ at which  the viscosity
 has decayed by $5\%$ from the maximum, {\it i.e.}, $\nu=0.95\,\nu_0$, while the second addresses  $\theta_2=\theta_1+\Delta \theta$ with  temperature increment $\Delta \theta=0.23$. The maximum viscosity decay rate always takes place at a constant temperature increment,  since the decaying rate of the law  (\ref{eqarcotang}) $\beta$,  is the same through out all this study.   At larger Rayleigh numbers,  $\Ra=3500$,  Figure \ref{b30b} shows that the head of the plume becomes more prominent. A comparison between   Figure \ref{b30b} and
  Figure \ref{b30c} indicates that the large viscosity contrast favors the formation of a balloon-shaped plume, with a thinner tail and more prominent and rounded head.    As regards the velocity fields, none of these patterns develop { a stagnant lid at the surface for any of the viscosity contrasts $a$ considered, even though the upper part corresponds to the region with maximum viscosity. This result is dissimilar to what is obtained in \cite{ULCRT12,CM13}.  In   \cite{CM13} it is argued that the cause of these differences could be attributed to the transition sharpness  controled by $\beta$, which in this work has been considered to be smoother. Additionally, the results reported in   \cite{ULCRT12} are obtained at larger viscosity contrasts, and the fact that these need to be large enough for the development of a stagnant lid has been addressed. }

%   \begin{figure*}
% %  \centering
%   \includegraphics[width=4.5cm]{b30a01R2500.pdf}  \includegraphics[width=4.5cm]{b30a01R3500.pdf}\includegraphics[width=4.5cm]{b30a0001R3500.pdf}
%    \caption{ Plumes obtained for the viscosity parameter  $b=30$. a) $R=2500$ and $a=0.1$; b)  $R=3500$ and $a=0.1$; c)  $R=3500$ and $a=0.001$.}\label{b30sa}
%   \end{figure*}
%     \begin{figure*}
%   \includegraphics[width=4.5cm]{b10a01R2500.pdf}  \includegraphics[width=4.5cm]{b10a01R3500.pdf}\includegraphics[width=4.5cm]{b10a0001R2500.pdf}
%    \caption{ Plumes obtained for the viscosity parameter  $b=10$. a) $R=2500$ and $a=0.1$; b)  $R=3500$ and $a=0.1$; c)  $R=2500$ and $a=0.001$.}\label{b10sa}
%   \end{figure*}
%      \begin{figure*}
%   \includegraphics[width=4.5cm]{b17a01R2500.pdf}  \includegraphics[width=4.5cm]{b17a0001R2500.pdf}\includegraphics[width=4.5cm]{b17a0001R3000.pdf}
%    \caption{ Plumes obtained for the viscosity parameter  $b=17$. a) $R=2500$ and $a=0.1$; b)  $R=2500$ and $a=0.001$; c)  $R=3000$ and $a=0.001$.}\label{b17sa}
%   \end{figure*}

We now consider  that  the parameter $b$ is small. As explained in Section 3, in this case the viscosity
transition occurs at low $\Ra$ numbers, below the instability threshold of the fluid
with constant viscosity $\nu_0$.
As low viscosity also implies diminishing the critical $\Ra$ number, the overall effect is that for small $b$
the instability threshold is below that with constant viscosity $\nu_0$, %Figure \ref{F2} confirms this for $b=10$.
and the phase transition  is perceived by weakly convective states. Figure  \ref{b10a} shows the structure of the plume obtained for $b=10$ and $a=0.1$ at $\Ra=2500$.
The head tends to be spread over a wide area and the viscosity transition occurs at  cold fluid zones away from the main plume.
This pattern is rather similar to those obtained with $b=5$ or $b=1$, except that  for smaller $b$ values the tail of the plume tends to be thinner.
Increasing the $\Ra$ number makes the tail of the plume thinner and  spreads  the head of the plume in the upper part, as reflected in Figure \ref{b10b}. On the other hand, high $\Ra$ numbers shift the viscosity transition  towards colder temperature contours. As expected from  the viscosity law (\ref{eqarcotang}), there is no $\Ra$ number at which the whole fluid layer is ``melted", since this law always imposes  that  a transition occurs  across the fluid layer. Figure \ref{b10c} reports the effect of diminishing the viscosity contrast $a$ to $a=0.001$ at $\Ra=2500$.  A mushroom-shaped plume with a thin tail and prominent head is observed. As before, none of these solutions develop a stagnant lid at the surface for any of the examined viscosity contrasts $a$.

 Intermediate values such as $b=17$  interpolate these extreme patterns. Figure \ref{b17a} shows the evolution from  Figure \ref{b10a} to  Figure  \ref{b30a} in which
 the black contour lines indicating the position of maximum viscosity decay converge towards the ascending  plume boundary, thus highlighting its shape. The head of the plume shrinks and the tail  strengthens.  Diminishing $a$ to the contrast 0.001 transforms the structure into a balloon-shaped plume  (Figure \ref{b17b}), while an increase in the $\Ra$
 number spreads the head of the plume in the upper fluid towards a mushroom-shaped plume.

 {The structure of the observed plumes as a spout, balloon or mushroom shape follows the schematic profiles reported in \cite{KK97}.
In the limit of low $b$, our viscosity law --as reported in Section 3-- converges towards the Arrehnius law used   by these authors, and the plume shapes reported there are similar to ours. However, a detailed comparison between both works is not possible as unlike these authors we  include the $\Ra$ number in the
 viscosity law, since this provides a  better expression of the realistic situation in which the increment of the $\Ra$ number is performed by increasing the
temperature differences between the bottom and upper surfaces.  Other viscosity laws, such as the exponential law reported in \cite{CuMa11} provide different plume structures, which are mainly spout shaped.}

The results reported in this section are obtained with expansions  {$(L\times M=37\times 44)$ except that in Figure \ref{b30c}, which corresponds to $(L\times M=47\times42)$.}
Similarly to what is reported in \cite{CuMa11}. The validity of these expansions is decided by ensuring that it provides accuracy in the eigenvalue along the neutral direction due to the SO(2) symmetry, which is always 0.  {This eigenvalue  is lost  if the expansions employed are insufficiently large, because badly resolved basic states present noisy structures either at the fields themselves or at  their derivatives,  and both contribute to the stability problem (\ref{eqtem23_2})-(\ref{eqtem43_2}). }

\subsection{Bifurcation diagrams and time dependent solutions}\label{SBD}
\begin{figure*}
   \centering
   \includegraphics[height=9cm]{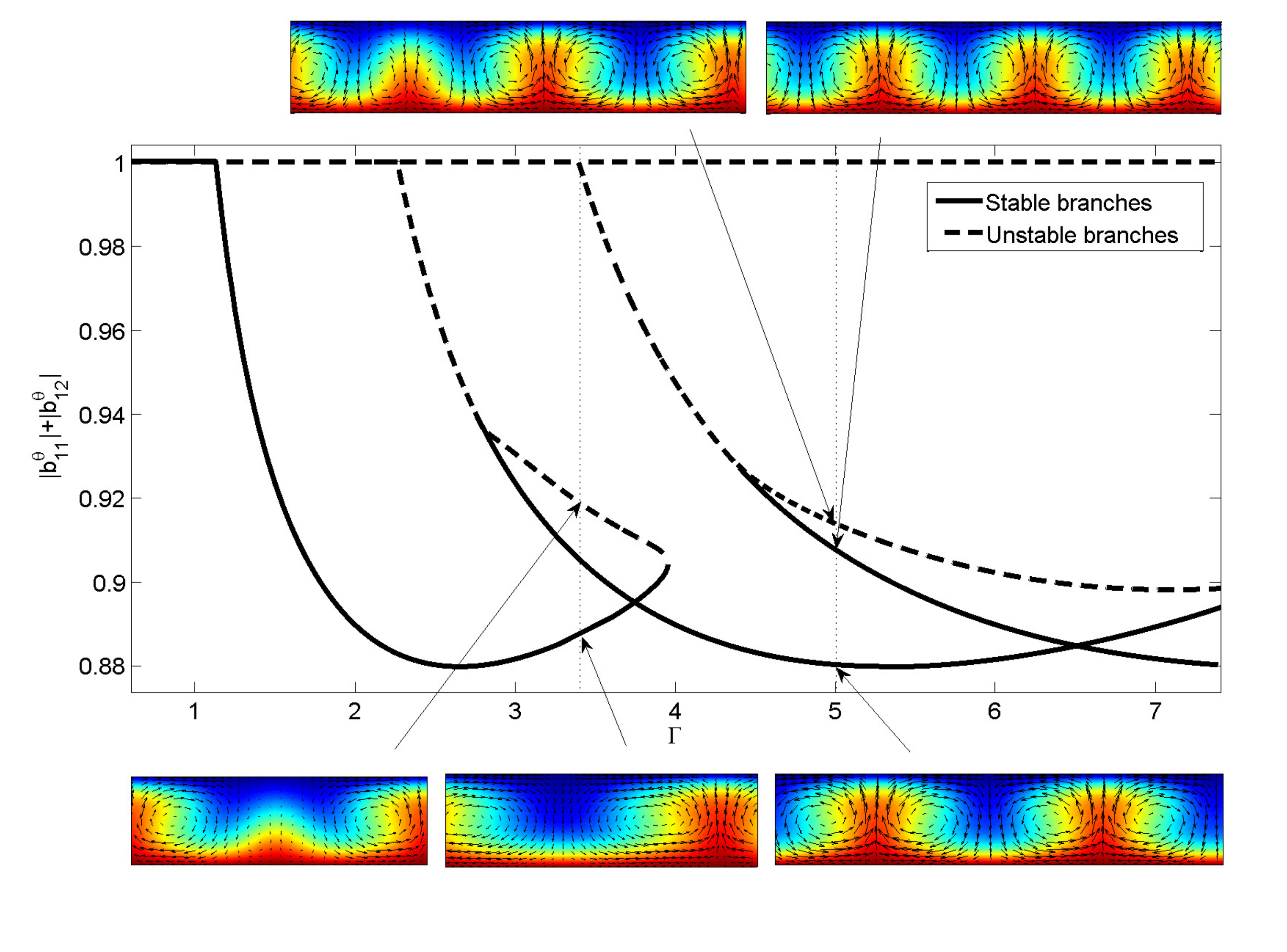}
    \caption{ (Color
 online). Bifurcation diagram as a function of the aspect ratio at $\Ra=1300$  for a fluid with viscosity dependent on temperature ($b=10$, $a=0.1$).  {Stationary solutions are displayed at different $\Ra$ numbers, which are highlighted by vertical lines. The arrows tag the branch points corresponding to the disclosed patterns. The dashed branches are unstable, while the solid ones are stable.}}\label{rama_asprat}
   \end{figure*}

Solutions to the system (\ref{eqproblem1})-(\ref{eqproblem3}) experience bifurcations  depending on the aspect ratio and on the Rayleigh number.
We now describe how these solutions vary along the dotted lines enhanced in Figure \ref{F2} for  parameters $\mu=0.0146$ and $b=10$. We consider for
 $a$ the choices 0.1 and 0.01.

Figure \ref{rama_asprat} shows the branch bifurcation diagram as a function of the aspect ratio for Rayleigh number $\Ra=1300$ and $a=0.1$.  Branches are obtained by representing along the vertical axis the sum of the absolute value of two relevant coefficients in the expansion of the temperature field, $b_{11}^{\theta}$ and $b_{12}^{\theta}$. Solid lines stand for stable branches, while dashed lines are the unstable ones.  {The horizontal line at $|b_{11}^{\theta}|+|b_{12}^{\theta}|=1$ corresponds to the trivial conductive solution.}   At a low aspect ratio, the stable branch is that with wave number  $m=1$, and at a higher aspect ratio  the stable solutions increase their wave number to $m=2$ and $m=3$. The unstable branch ending up with a saddle-node bifurcation and connecting the $m=1$ with the $m=2$ branch corresponds to a mixed mode.

Stationary stable and unstable solutions, obtained at the positions indicated by arrows, are pictured.
%In the patterns, the two black contour lines mark temperatures between which the viscosity  decays  most rapidly.
No stagnant lid appears at the surface for any of the  aspect ratios considered. The expansion orders required  by this figure to ensure accuracy are not the same along all branches.  {We have guaranteed that for successive orders expansions the amplitude values displayed on the vertical axis of the  bifurcation diagrams   are preserved.}
A rule of thumb is that high modes obtained at larger aspect ratios require higher expansions.
Thus while for mode $m=1$ expansions $(L\times M=37\times 44)$ are sufficient, for $m=2$ and $m=3$ at larger aspect ratios expansions are increased up to $(L \times M=61\times  44)$.

\begin{figure*}
   %\centering
   \includegraphics[height=9cm]{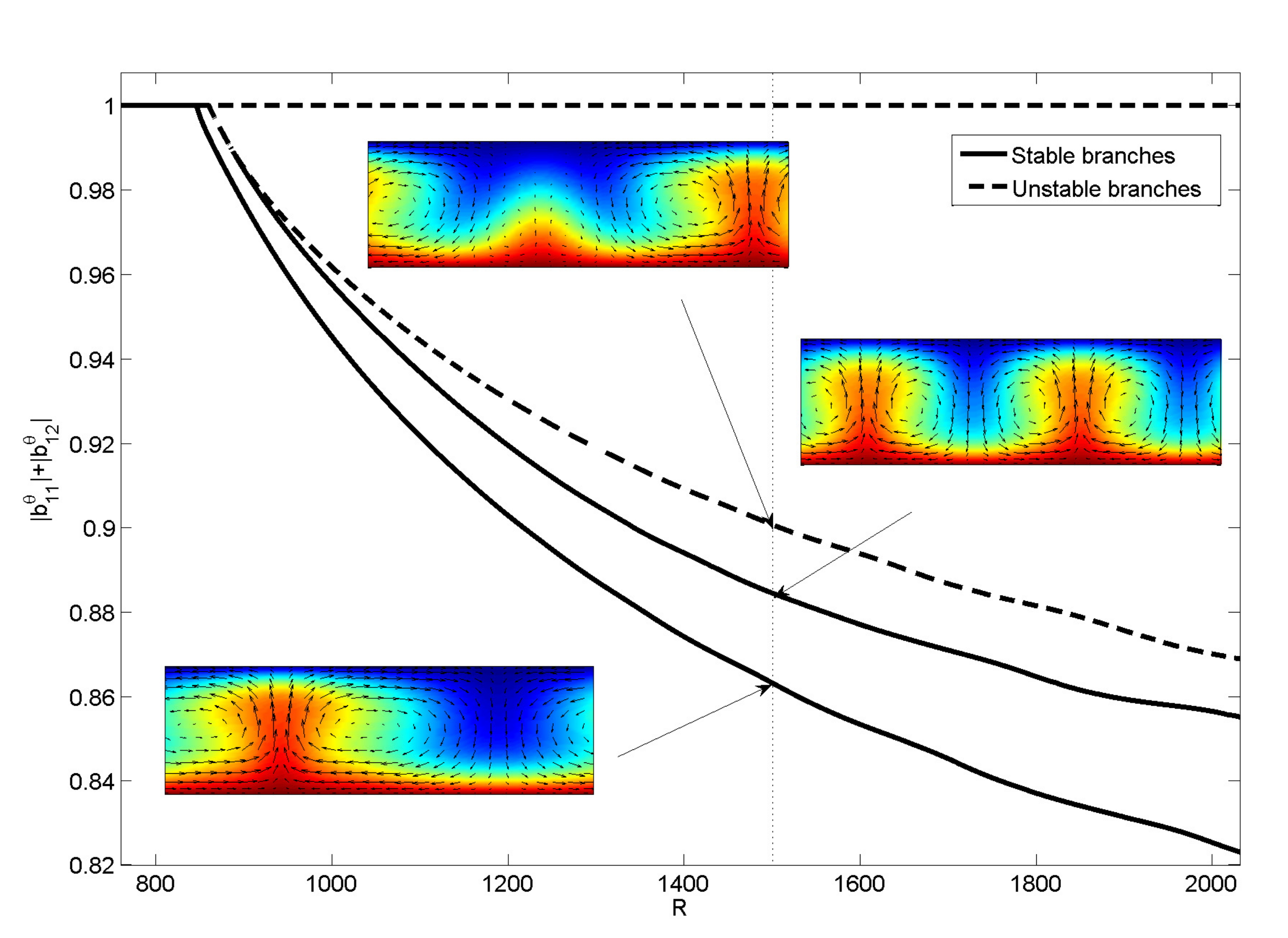}
    \caption{ (Color online). Bifurcation diagram as a function of the Rayleigh number for a fluid with viscosity dependent on temperature ($b=10$, $a=0.1$) at $\Gamma=3.4.$  {Stationary solutions are displayed at the $\Ra$ number, which is highlighted with the vertical line. The arrows tag the branch points corresponding to the disclosed patterns.The dashed branches are unstable, while the solid ones are stable.}}\label{RamasG3_4}
   \end{figure*}

Bifurcations are further analyzed  at three different aspect ratios as a function of the Rayleigh number.  {Among the many possible choices   for the aspect ratios,  we consider  occurrences  at which the existence of solutions related to symmetries are found, such choices  thereby serving  our purpose of highlighting  the importance of symmetries in fluids with viscosity dependent on temperature}.  Figure \ref{RamasG3_4} represents the branching obtained at
$\Gamma=3.4$ for $a=0.1$.
The pictured plumes, which are  computed for  a rather low Rayleigh number, $\Ra=1500$,  are  spout-shaped, with the tail of the plume  nearly as large as the
head. As already reported in the previous section for increasing $\Ra$ numbers, plumes become balloon-shaped and beyond that mushroom-shaped. No stagnant lid is observed at any $\Ra$ number.
%The black contour lines highlight the temperatures between which the viscosity  decays more rapidly.
Several branches are distinguished.
The branch related to mode $m=1$ arises at the lowest $\Ra$ number and is stable in the whole range displayed. Mode $m=2$ emerges at $\Ra\sim 860$ from the unstable conductive solution through an unstable branch, which becomes stable through a pitchfork bifurcation at $\Ra\sim 890$. Results at this aspect ratio  are obtained
with expansions $(L\times M=37\times 44)$.

\begin{figure*}
   \centering
   \subfigure[]{\includegraphics[height=9cm]{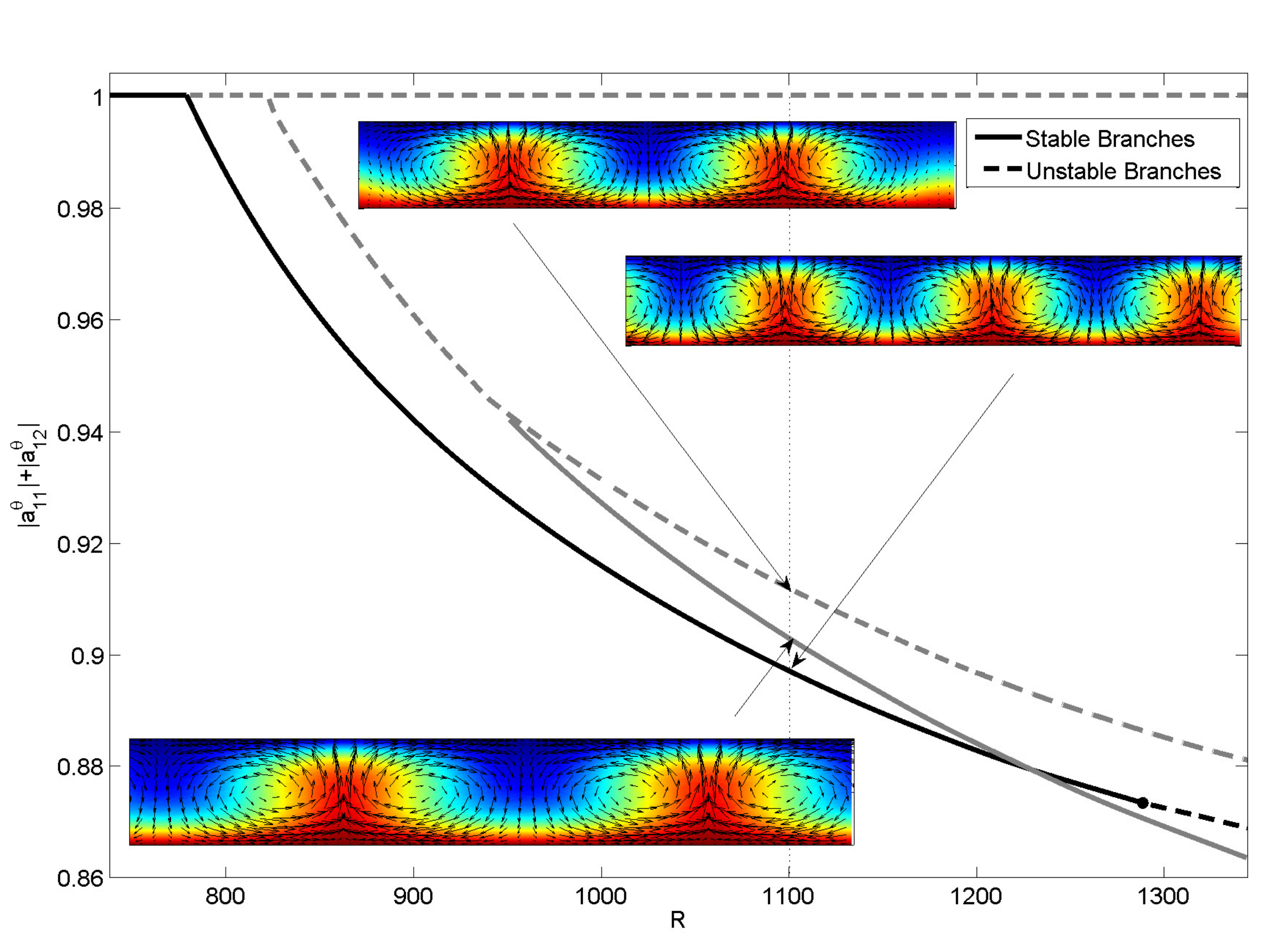} \label{RamasG6_9a}}
   \subfigure[]{\includegraphics[height=9cm]{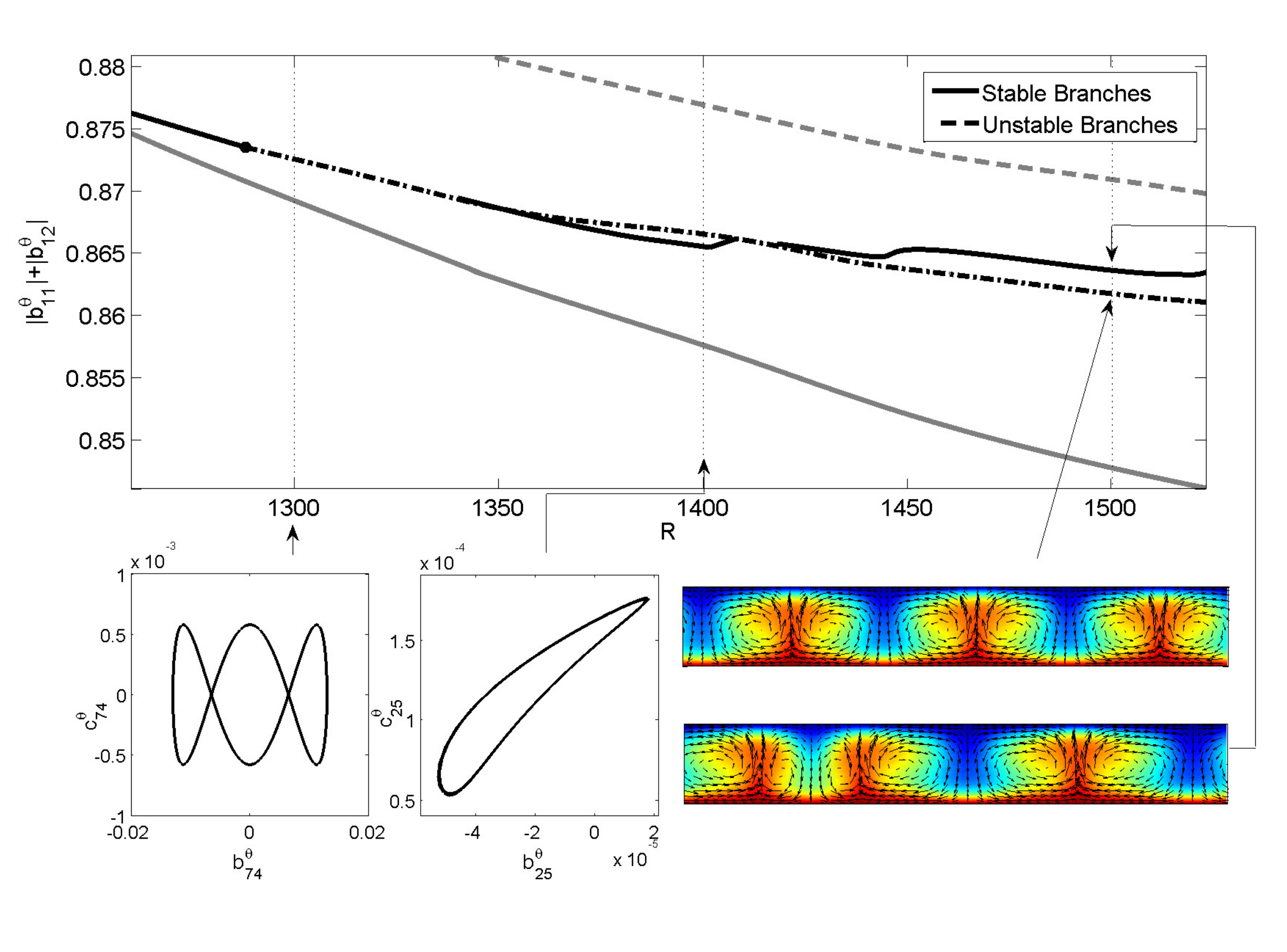}\label{RamasG6_9b}}
    \caption{(Color online). Bifurcation diagrams as a function of the Rayleigh number for a fluid with viscosity dependent on temperature ($b=10$, $a=0.01$) at $\Gamma=6.9.$  {The dashed branches correspond to stationary unstable solutions, while solid branches correspond to stationary stable ones. The gray lines indicate   spatial patterns with period 2, while the black ones are for  period 3 patterns.} a) Rayleigh number in the range 800-1300.  {Stationary solutions are displayed at the $\Ra$ number highlighted with the vertical line. Arrows tag the branch points corresponding to the disclosed patterns}; b) Rayleigh number in the range 1250-1500.  {Stationary solutions are displayed at the $\Ra$ number, which is highlighted by the vertical line.  The arrows tag the branch points corresponding to the disclosed patterns.  Two additional vertical lines highlight the $\Ra=1300 $ and $\Ra=1400 $ numbers at which time dependent solutions are found. These are displayed as a time series projected on the coefficient space (for a description see the text and \cite{supplemental}).}}\label{RamasG6_9}
   \end{figure*}

This simple diagram with simple stationary solutions obtained at a low aspect ratio is in contrast to  those
with more complex solutions obtained at a larger aspect ratio.  Figure \ref{RamasG6_9} represents  the bifurcations obtained at $\Gamma=6.9$ as a function of $\Ra$ for $a=0.01$.
   Figure \ref{RamasG6_9a} examines the $\Ra$ interval from 800 to 1300. In this range several stationary solutions are portrayed both stable and unstable.
   %As in previous diagrams  black contour lines mark the temperatures at which  the viscosity has the largest decaying rate.
    At $\Ra\sim 1290$, a Hopf bifurcation occurs at the branch of mode $m=3$  {(see Figure \ref{RamasG6_9b})}.  After the bifurcation, a traveling wave is found, as   illustrated in the phase portrait represented at $\Ra=1300$. The solution evolves in time by traveling  towards the left.   This  breaks the symmetry $x \to -x$. However, the right traveling  solution obtained by the symmetry transformation also exists, as expected from equivariant bifurcation theory \cite{CK91}. { See \cite{supplemental} for further details}. The presence of traveling waves after a Hopf bifurcation has been reported in diverse contexts in
   under the presence of the O(2)  symmetry \cite{AGH88,vGMP86,PoMa97,CK91}, and here they are  reported   in the context of convection with variable viscosity.
  At larger $\Ra$ numbers, up to $\Ra\sim 1320$, the traveling wave persists, while its frequency increases.  A stable fixed  point with wavenumber $m=3$ is found in the range $\Ra\sim 1340-1380$. A cycle limit appears at around $\Ra\sim 1400$. In this regime,  the time-dependent solution consists of plumes that weakly oscillate in the horizontal direction around their vertical axis of symmetry. {See \cite{supplemental} for further details}. Close to $\Ra\sim 1416$, a stable branch of fixed points emerges, which is visualized at $\Ra\sim 1525$. It shows the presence of  plumes that are non-uniformly distributed along the horizontal coordinate:  two close plumes, which are asymmetric around  their vertical axis, and a third one  that maintains its symmetry.  None of the described solutions develop stagnant lids at the surface. At low $\Ra$ numbers  (\emph{i.e.} Figure \ref{RamasG6_9a}) results are obtained with expansions $(L\times M=47 \times 44),$ while for higher  $\Ra$ numbers  (\emph{i.e.} Figure \ref{RamasG6_9b}) results are obtained with
 expansions $
 (L\times M=61 \times 44)$.

   \begin{figure*}
   \centering
   \subfigure[]{\includegraphics[height=9cm]{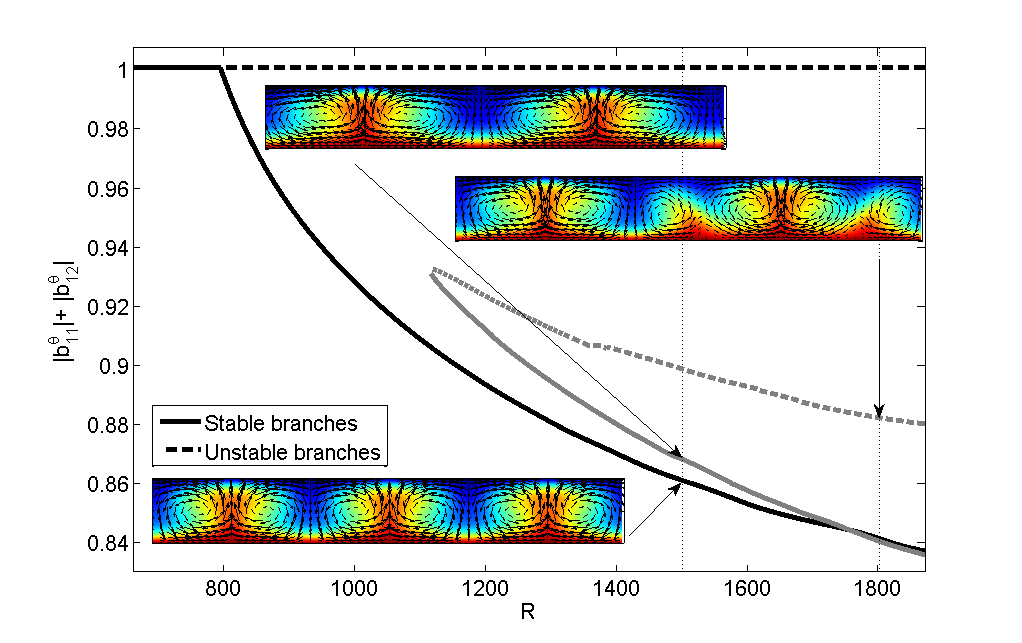}\label{RamasG7_4a}}
   \subfigure[]{\includegraphics[height=9cm]{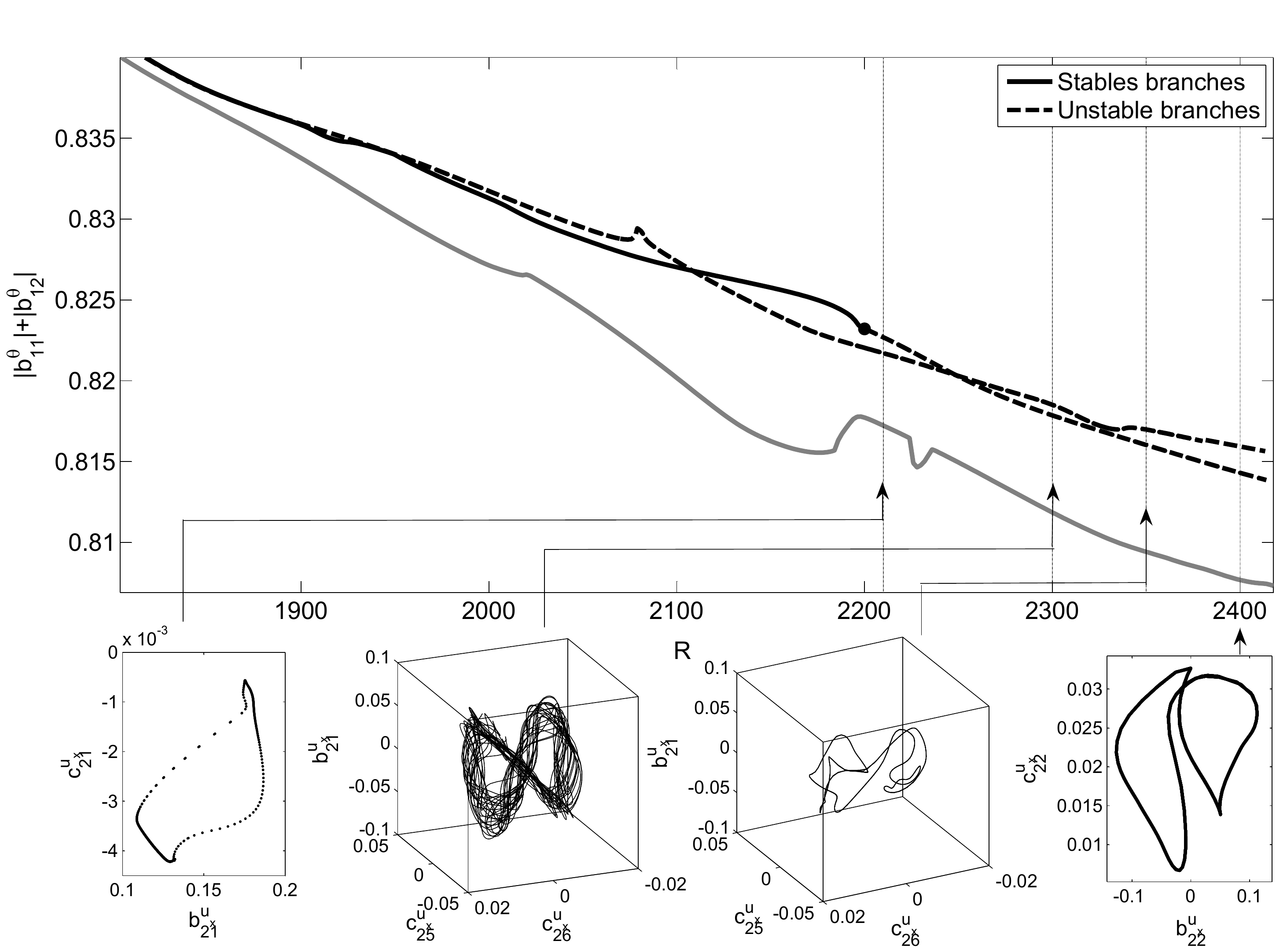}\label{RamasG7_4b}}
    \caption{(Color online). Bifurcation diagrams as a function of the Rayleigh number for a fluid with viscosity dependent on temperature ($b=10$, $a=0.1$) at $\Gamma=7.4.$  {The dashed branches correspond to stationary unstable solutions, while  solid branches correspond to stationary stable ones. The gray lines stand for  spatial patterns with period 2 while the black ones are for  period 3 patterns.} a) Rayleigh number in the range 700-1800.  {Stationary solutions are displayed at the $\Ra$ number, which is highlighted by the vertical line.  The arrows tag the branch points corresponding to the disclosed patterns}; b) Rayleigh number in the range 1800-2500.  {Vertical lines highlight the $\Ra$ numbers at which time dependent solutions are found. These are 2210,  2300, 2350 and 2400. These are displayed as a time series projected on the coefficients space (for  a description see the text and  \cite{supplemental}). }}\label{RamasG7_4}
   \end{figure*}

Figure \ref{RamasG7_4}  shows the bifurcation diagram obtained at $\Gamma=7.4$ as a function of $\Ra$ for $a=0.1$.
The mode $m=3$ branch,  marked with a solid black line, emerges at $\Ra\sim 794$.  { Figure \ref{RamasG7_4b}  shows that at}  $R\sim 2190$
the branch undergoes a Hopf bifurcation. Beyond this point, solutions embedded in a projection over the coefficient space are represented at the $R$ values marked with vertical dotted lines. A limit cycle is observed at $\Ra=2210$ just above the bifurcation point. Its projection over the coefficient space displays a point at every time step of the time series.
The solution  appears to reside  in the neighbourhood of a  heteroclinic connection between two fixed points  as it evolves into  a quasi-stationary regime --near the large density of points-- followed by a rapid transition to a new quasi-stationary regime. The two fixed points between which the solution oscillates are similar to the non-uniformly distributed plumes described in the previous paragraph {(see \cite{supplemental} for further details).   A solution is found at $\Ra=2300$ that has a time-dependence in which the block of  plumes  shifts irregularly along the horizontal direction, towards both the left and the right (see \cite{supplemental}). For increasing
$\Ra$ numbers, the horizontal motion persists, but the oscillation  becomes more regular and pattern displacements along the $x-$coordinate are gradually reduced. This is verified
through simulations at $\Ra=2350$ and at $\Ra=2400$ (see \cite{supplemental}).} The diagram  displayed in Figure \ref{RamasG7_4a} shows a gray solid line  associated to a mode $m=2$ stable branch that emerges by means of a saddle node bifurcation jointly with an unstable branch. An irregular pattern  obtained at $\Ra=1800$ for the unstable branch is included in this diagram.
Once again, none of the solutions described at this aspect ratio has a stagnant lid at the surface. Results in this figure are obtained with different order expansions. At low $R$ number expansions  {$(L\times M =47 \times 42)$ are sufficient while for higher $\Ra$ numbers they are increased up to $(L\times M=61 \times 44)$  and even to $(L\times M=101 \times 44)$.}

{The time dependent solutions reported in Figures 9 and 10 in many respects resemble  those  described for the Kuramoto-Sivashinsky (KS) equation \cite{PoMa97,HNZ86} in the presence of the O(2) symmetry, which also report the presence of traveling waves and heteroclinic cycles.
The KS equation is proposed in order to describe thermal diffusive instabilities in flame fronts \cite{S80}, and while apparently this setting is rather different to ours, the similitude between  solutions suggest that the abrupt changes in the viscosity could define a similar kind of front to those observed in  flame propagation phenomena. On the other hand, similar solutions have been found in 3D convection
with constant viscosity in the presence of the O(2) symmetry \cite{EKG97,DCJ00}, thus confirming the determining role of the symmetry  in the dynamics. }

%\section{Solutions dependents of time}\label{Sdt}
 \section{Conclusions}
 \label{S5}
This article addresses the study of a convection problem with temperature-dependent viscosity in the presence of the O(2) symmetry. In particular, the
considered viscosity law represents a viscosity transition at a certain temperature interval around a temperature of transition. This is a problem of great interest for its many applications in geophysical and industrial flows  {and in this work the  focus is on exploring the impact of symmetry on the solutions displayed by system}.

Our results report  the influence on parameters $a$ and $b$ of the viscosity law on the morphology of the plumes at a low aspect ratio ($\Gamma=3.4$).  It is shown that if the temperature of transition is well above  the instability threshold of a fluid  with constant viscosity $\nu_0$, {\it i.e}, $b$ is large, plumes tend to be thicker and  show
spout-like shapes. Increasing the $\Ra$ number induces their evolution towards balloon-shaped plumes, and this effect is more pronounced for high viscosity contrasts (small $a$).
At low $b$ values plumes are thinner, and the head of the plume tends to spread in a mushroom-like shape in the upper part of the fluid.

We explore bifurcations  both for a fixed $\Ra$ number as a function of the aspect ratio, and bifurcations at three fixed aspect ratios as a function of the $\Ra$ number. No stagnant lid regime is observed in any of the physical conditions analyzed.
Among the stationary solutions obtained along the bifurcation branches, one of the more interesting stable patterns consists of the
non-uniformly distributed plumes that break symmetry along their vertical axis.

We also find  that, for the higher Rayleigh numbers explored, at a high aspect ratio several rich dynamics appear.
As already reported in classical convection problems, %\cite{CK91,GS85,GSS00}.
 we find  dynamical phenomena fundamentally related
to the presence of symmetry, such as traveling waves, oscillating  solutions in the neighborhood of heteroclinic connections and chaotic  regimes characterized by  ``phase" drifts along
the horizontal direction linked to  the SO(2) symmetry.

 \section*{Acknowledgements}
We are grateful to CESGA and to CCC of Universidad Aut\'onoma de Madrid for computing facilities.
This research is supported by the Spanish Ministry of Science under grants  MTM2008-03754, MTM2011-26696 and  MINECO: ICMAT Severo Ochoa project SEV-2011-0087.
\bibliographystyle{plain}
\bibliography{local}

\end{document}